\documentstyle[prb,aps,eqsecnum]{revtex}
\begin{document}
\draft
\title{Periodic Molecular Arrangements from Space Groups.} 
\author{  Ravi Sachidanandam$^{1,2}$ and A. B. Harris$^1$}
\address{ 
	1. Department of Physics, 
         University of Pennsylvania, 
         Philadelphia, PA 19104. }
\address{2. School of Physics and Astronomy, Raymond and Beverly Sackler
Faculty of Exact Sciences, \\ Tel Aviv University, Tel Aviv 69978, Isarel}
\maketitle
\begin{abstract}
We demonstrate, using examples from $2$ and $3$-dimensions, a
systematic method of finding all possible periodic arrangements of a
given molecule or molecules such that the arrangements have the
symmetry of a given space group. The technique is based on work by
Rabson et al.  We give tables of generators which facilitate such work.
\end{abstract}
\pacs{ 61.10.-i; 61.50.Em; 64.70.Kb}

\input{psfig}

\section{Introduction}
The problem of determining all possible periodic arrangements of
molecules within a give space group frequently arises in the study
of structure using  X-ray diffraction and elastic neutron scattering.
The resolution of this issue needs a thorough understanding of space
groups. 

A new method of deriving and listing the crystallographic
space groups, introduced by Rabson {\it et.\ al.\ } and Mermin (RMRW)
\cite{MERMIN}  has simplified the derivation of space
groups and made compact presentation possible. This has also
unified the subjects of crystallography and quasicrystallography
but, unfortunately, their techniques  have not yet found
widespread acceptance with researchers  studying periodic
structures in condensed matter physics. One  reason for this
may be that RMRW work in Fourier space while the arrangement of
molecules is easier to visualize in real space. We will
restate the algebraic relations used by RMRW in real space
language and use their technique to list the crystallographic
space groups compactly in real space.

We will then demonstrate
our technique of using these algebraic relations to 
find molecular arrangements from these listings. 
Typically, the orientation of molecules at low
temperature in a molecular crystal is determined by
elastic diffraction, either of X-rays or of neutrons.
If a large amount of high quality data is available,
there may not be a problem in determining the crystal
structure using standard computer programs to carry out
the refinement.  However, a more delicate approach,
 one that we used with reference to C$_{60}$\cite{SHPRL,HSPRB}
and which also has been used for other fullerene systems\cite{TYPRL}
is the following.  For these systems, the orientationally
disordered phase has been determined to be face-centered cubic (FCC). 
At the ordering temperature, new superlattice reflections appear,\cite{PAHPRL}
indicating that the molecules develop long--range orientational
order.  From the outset,\cite{PAHPRL} it was clear that
the low temperature diffraction pattern had to be indexed with
a simple cubic (SC) unit cell containing four orientationally inequivalent
molecules.  There was no suggestion that the
centers of mass were no longer on an FCC lattice.  To
narrow down the possible structures, it was therefore
sensible to ask the question: how can one arrange icosahedra
on an FCC lattice such that the SC unit cell contains four
molecules whose orientations are fixed?\cite{ERRATUM}  The
determination of 
the allowed structures, led to an easy
identification\cite{SHPRL} of the space group of C$_{60}$.
This structure has been identified by several
groups.\cite{REPLY,WIFD,MICH}
More recently, with the suggestion\cite{2a} of a complicated
("2--$a$") structure for $C_{60}$ (based on a still larger
unit cell), a variant of the question arose,\cite{2aPRL,OUR2a} namely,
how to arrange icosahedra on FCC lattice sites such
that there are eight molecules per FCC unit cell.
(In this case, the FCC unit cell would have a lattice constant
twice as large as that of the orientationally disordered lattice.)
However, it is now generally agreed\cite{AGREE} that such a structure
is not an equilibrium phase of C$_{60}$.  The method we
used\cite{HSPRB,OUR2a} in this work was, inelegant, however.
It involved an explicit search through the listings of
the crystallographic space groups in the International Tables
of Crystallography (ITC).\cite{ITC}  The symmetries of all points
in the unit cell are listed there, and, in principle, this
information suffices to identify all allowed structures.  However,
a) such a direct consideration of all possibilities is tedious,
b) in such an enumeration it is easy to overlook possible
allowed space groups,\cite{ERRATUM} and c) the tables themselves
are not very easy to use without some background in crystallography.
One advantage of the technique given here is that it is
completely algebraic and the algorithm is a constructive one,
rather than one of exhaustive enumeration.  Also the method 
requires only an understanding of the symmetries of simple
objects like the square, the cube, etc.  In the course of
this work, we have also become aware of some confusion concerning
the proper formulation of this problem, associated with
the question of what molecular distortions should or should not
be considered in determining the allowed structures of
molecular crystals.

This paper is organized as follows. In Sec. \ref{SG} we define some
terms and show how space groups can be specified  compactly. In
Sec. \ref{rel} we state the relations we will be using. In Sec.
\ref{prop} we point out some basic properties that will be used in our
technique. In Sec. \ref{tech} we illustrate our technique by
considering a hypothetical question concerning the arrangement of
hexagonal (benzene) and octagonal molecule on a square lattice.
In Sec. \ref{3d} we apply the technique to
problem concerning $3$-dimensions molecular crystal,
in particular those of H$_2$ and C$_{60}$.  A brief summary
of our work is given in Sec. \ref{SUMMARY}.  In the course
of this work it is helpful to relist all the cubic
space groups in the RMRW language.  (The similar list extended
to all noncubic space groups is also available.\cite{DEPOSIT})

\section{Space Groups}
\label{SG}

In this and the next two sections we briefly review some of the
basic facts concerning space groups and their characterization.
In this we adapt the ideas of RMRW.

The set of  symmetries of a periodic array is the space group
of that array.  There are only a finite number of distinct space
groups ($17$ in $2$-dimensions, $230$ in $3$-dimensions). We will
see that to  specify a space group it is necessary to specify the
lattice, the point group, and the  translations associated with the
generators of the point group. The meaning of these terms will be
clarified in the following paragraph. 

In $2$-dimensions, a {\bf Bravais lattice} (henceforth just a  
lattice) is the set of points $\{ {\bf r} \}$ defined by, 
\begin{equation}
{\bf r}  = n_1 {\bf a}_1 + n_2 {\bf a}_2  , 
\end{equation}
where ${\bf a}_1, {\bf  a}_2 $ comprise  a linearly independent
set of basis vectors and the $n_i$ are integers.
Such a vector ${\bf r}$ is called a lattice vector.
In $d$-dimensions there is a linearly independent set of $d$
basis vectors. From the definition it is obvious that all
lattices have inversion symmetry ($n_i \rightarrow -n_i$).
The choice of basis vectors is not unique since any linearly
independent set constructed from the ${\bf a}_i$ can 
equally well be chosen as basis vectors. The obvious symmetries
of any  lattice are the translations by lattice vectors.
The {\bf nontranslational} symmetries, consisting of 
rotations, mirror planes, and combinations thereof,
are the ones that specify a lattice uniquely.   There are
only a finite number of lattices with distinct symmetry,
five in $2$-dimensions and fourteen in $3$ dimensions.
For example, the square lattice of   Fig.\ \ref{sq}
is one of the five lattices in $2$-d. 
Each of its points has the following symmetry elements
in which the points itself remains fixed: $M$, a mirror
plane parallel to a crystallographic $(1,0)$ direction,
$M'$, a mirror plane parallel to a diagonal, and $R_4$,
a $4$-fold rotation about an axis perpendicular to the
plane.  There are a total of eight elements in the symmetry
group of the point, $E$, $M$, $R_4$, $M'=R_4 M$,
$R_4MR_4^{-1}$, $R_4M'R_4^{-1}$, $R_4^2$ and $R_4^3$, where
$E$ is the identity and $R_4^n$ means a $4$-fold rotation is
applied $n$ times.  Any symmetry group can be completely
specified by a few elements of the group called the
{\bf generators}.  For example, the symmetry group of the
square, denoted $4mm$, is completely specified by specifying the
existence of the generators $R_4,M$ and
the generating relations $R_4^4 = M^2 =E, (R_4 M)^2 = E$. 

In this paper we use the following conventions.  Vectors are 
always specified by their components along the lattice vectors. 
With the symbols for vectors, a  {\bf superscript}  
represents the component of the vector and a {\bf subscript} 
identifies  a particular  vector. For example, $t_g^1$ is the 
component of vector ${\bf t}_g$ along the basis vector
${\bf a}_1$. A {\bf superscript} on the symbol for a symmetry
element represents the number of times the operation is carried out.
For example, applying the mirror $m$ twice to any structure  brings it
back to its original state, or $M^2 = E$.  The symbol $\equiv$ is
used to signify that the vectors on both sides of the symbol differ
from each other only by a lattice vector, for example
$\frac{1}{2} {\bf a}_1 \equiv \frac{3}{2} {\bf a}_1$. 
Finally, to avoid confusion with the operations
in three dimensions, we will
denote point group operations in two dimensions by capital
letters and those in three dimensions by small letters.

Any periodic structure can be considered to be a decoration of
a lattice and the obvious symmetries of such a structure are 
the  translations of the structure by a lattice vector. The
decorations preserve some of the non-translational symmetries
of the lattice unchanged, while a few of these symmetries,
of which say $g$ is an example, require an additional
{\bf non--lattice translation}, say ${\bf t}_g$, to remain as
symmetries. Some of the symmetries of the lattice may be completely
lost.  Specifying the unbroken symmetries of the lattice along
with the required accompanying non--lattice translation completely
specifies the symmetry of the space group. All the nontranslational 
symmetries of the lattice that are {\bf not} lost form the {\bf point
group} of the space group.  Obviously, if any ${\bf t}_g$ is
nonzero, the point group is not actually a subgroup of the
space group.  In the literature such space groups are called
nonsymmorphic.
 
Consider  the square lattice shown in Fig.\ \ref{sq}. Decorating
it as shown in Fig.\ \ref{sqspace} (a) leads to all the symmetry
elements of the lattice being preserved. Decorating the same square
lattice with the pattern shown in Fig.\ \ref{sqspace} (b) leads to 
the mirror $m$ requiring  the translation
${\bf t}_M = (\frac{1}{2},\frac{1}{2})$
to follow it to get back the original structure. Since none of
the symmetries are lost completely, the point group of both
structures is $4mm$, the symmetry of the square.  Decorating the
lattice as shown in Fig.\ \ref{sqspacec} leads to the loss of
mirror symmetries. The point group of this structure is now
the point group with only the $4$-fold rotations, $C_4$ or $4$.
These are the only distinct space groups with a square lattice.
Table \ref{t2dimage} lists the effect of the symmetry
elements of a square on the point $(x,y)$. Table \ref{tsqsg}
lists all the space groups with a square lattice and point group 
$4mm$. These data will be used later. 

\section{The Algebraic Relations.}
\label{rel}

We now show that specifying the translations associated with the
generators of the point group is enough to specify the translations
associated with all the other elements of the point group. 
If the symmetry of a periodic structure is of the form $(g,{\bf t}_g)$,
$(h,{\bf t}_h)$ etc., where $(g,{\bf t}_g)$ means the action of the 
element  $g$ is followed by a nonlattice  translation
${\bf t}_g$ and $g$, $h$ etc.
are the elements of the point group,  then it is easy to show (See
Appendix A) that 
\begin{equation}
\label{comp}
{\bf t}_{gh} \equiv g {\bf t}_h + {\bf t}_g,
\end{equation}
where ${\bf t}_{gh}$ is the translation associated with the
element $gh$.  (The operation $gh$ is the operation $h$ followed
by the operation $g$.)
This is called the {\bf  Compatibility relation}  or the  {\bf
Frobenius relation}\cite{FROB} In view of Eq.\
\ref{comp} knowledge of the translations associated with the
generators is enough to find the translations associated with all the
elements of the point group.
For example in Fig.\ \ref{sqspace} (b) we have
${\bf t}_{R_4} \equiv (0,0), {\bf t}_M \equiv (\frac{1}{2},\frac{1}{2})$.
Since $M' = R_4 M$ we get from Eq.\ \ref{comp}
${\bf t}_{M'} \equiv {\bf t}_{R_4 M} \equiv R_4 {\bf t}_M + {\bf t}_{R_4}
\equiv (\frac{1}{2},\frac{1}{2})$. 

\begin{figure}
\centerline{\psfig{figure=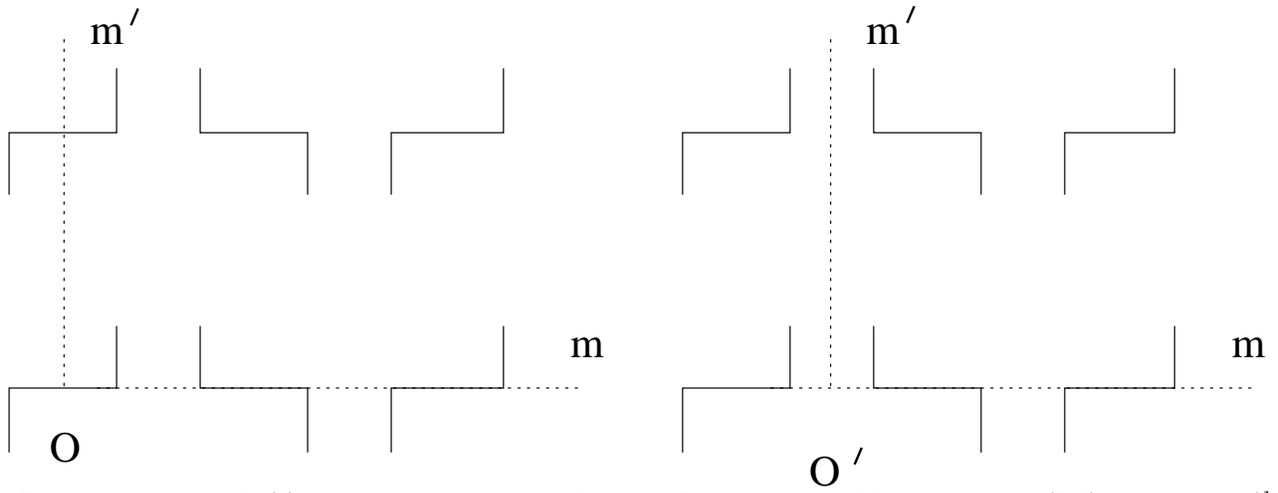}}
\caption{\label{origin}
The structure in (a) has  rectangular lattice, point group $2mm$. 
For  the origin chosen ${\bf t}_{R_2} \equiv (0,0)$,
$ {\bf t}_M \equiv {\bf t}_{M'} \equiv (\frac{1}{2},0)$.
(b) shows the same structure with a different choice of origin.
For the origin here,
${\bf t}_{R_2} \equiv (\frac{1}{2},0) \equiv  {\bf t}_M $,
$ {\bf t}_{M'} \equiv (0 , 0)$.}  
\end{figure}

\begin{figure}
\centerline{\psfig{figure=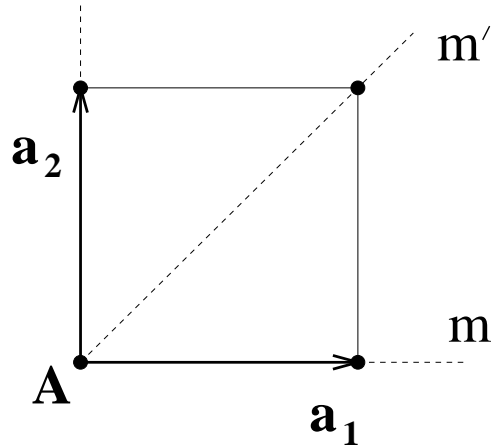,width=2.5in}}
\caption{The symmetries of a square.}
\label{sq}
\end{figure}

The symmetry axes of rotations and the mirror planes of a point group
contain a point designated as the origin,  which they leave invariant.
In the space group operation, $(g, {\bf t}_g)$, the translation,
${\bf t}_g$, associated with the element $g$ of the point group is
dependent on the choice, which is arbitrary, of the origin.
For example,  consider the structure shown in Fig.\ \ref{origin}.
Its point group is $2mm$, the point group of a rectangle. 
Let us consider the translational part of the space group
operations involving $R_2$, a two--fold rotation about an axis
perpendicular to the page and which passes through the
origin and that involving $M$, a horizontal mirror plane 
perpendicular to the plane of the paper passing through the origin.
When the origin is at 
point O, then ${\bf t}_{R_2} \equiv (0,0)$ and
${\bf t}_{M} \equiv (\frac{1}{2},0)$.  In contrast, when $O'$
is chosen as the origin then ${\bf t}_{R_2} \equiv (\frac{1}{2}, 0)$
and ${\bf t}_{M} \equiv (0,0)$. 
There exists a relationship between the translations associated with 
the elements with respect to the two  origins. We use
${\bf c} \equiv (c_1,c_2)$ to denote the coordinates of the new origin
with respect to the old one. The components of the vector are along
the basis vectors of the lattice.
When the origin is shifted by ${\bf c}$ then the
translations associated with the element $g$ change
by $\Delta {\bf t}_g$ which is obtained from the {\bf origin shift}
relation stated as, 
\begin{equation}
\label{oshift}
\Delta {\bf t}_g \equiv   g {\bf c} - {\bf c} .
\end{equation}
A proof of this relation is given in appendix \ref{aoshift}. Using the
two relations of this section it is easy to derive all the space
groups, as shown by  RMRW \cite{MERMIN}. 

\section{Useful Properties}
\label{prop}

If, for the origin chosen, the translations associated with
the elements $g_1,g_2 \ldots g_n$ are zero, that is,
${\bf t}_{g_1}$ $\equiv$
${\bf t}_{g_2}$ $\ldots {\bf t}_{g_n}$ $\equiv 0$
then the origin is said to have a
{\bf local} or {\bf site} symmetry  given by the group made up of 
$g_1,g_2 \ldots g_n$. Thus, the local symmetry of any site in the unit
cell can be found by translating the origin to that site and then
finding which symmetry elements have zero translations associated
with them.  This idea will be used frequently in this chapter. 

\begin{figure}
\centerline{\psfig{figure=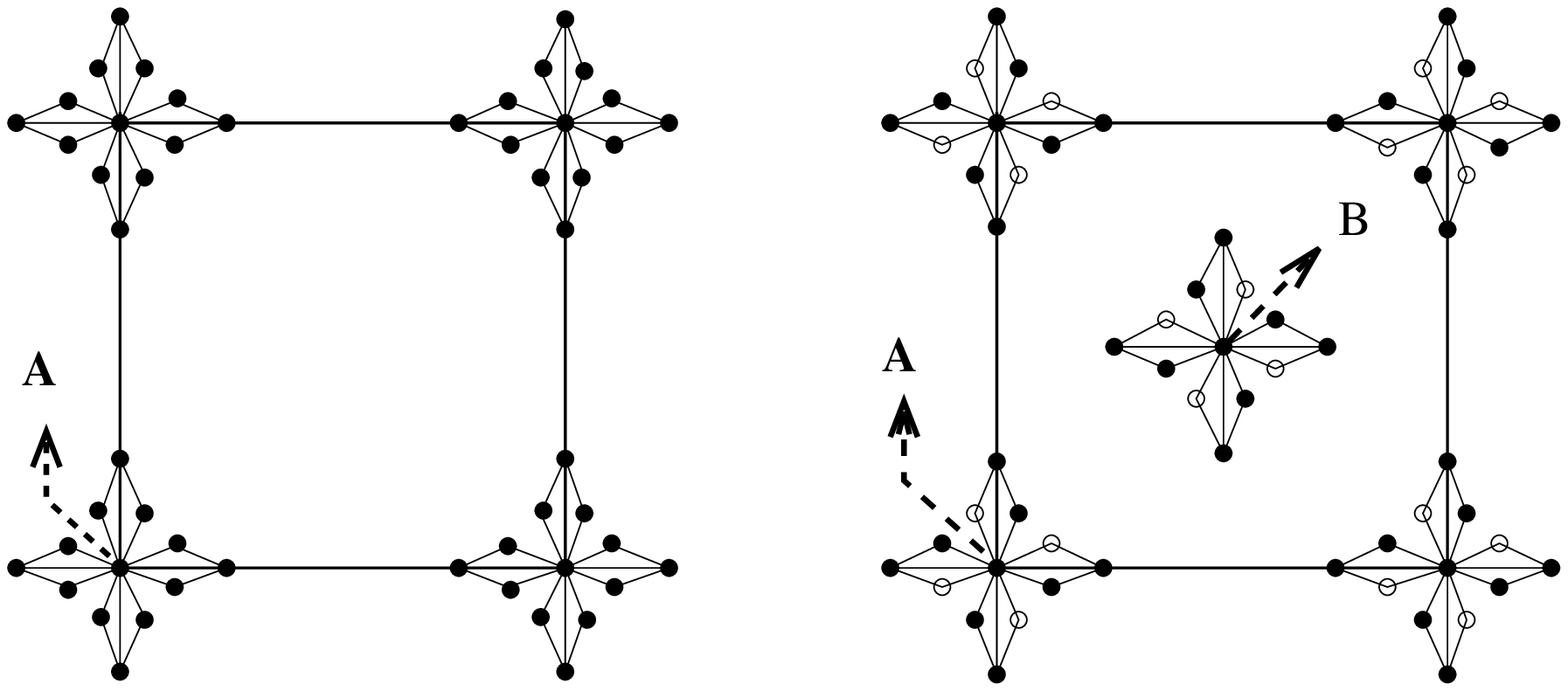}}
\caption{
\label{sqspace} (a) shows a decoration of the square 
lattice. The structure has  point group $4mm$ and  
$\protect{\bf t}_{R_4} \equiv \protect{\bf t}_M \equiv (0,0)$. 
(b) shows a decoration  of the  square
lattice with a symmetry distinct from (a). The structure has 
point group  $4mm$ and  $\protect{\bf t}_{R_4} \equiv (0,0)$,
$\protect{\bf t}_M \equiv (\frac{1}{2}, \frac{1}{2})$.}
\end{figure}

\begin{figure}
\centerline{\psfig{figure=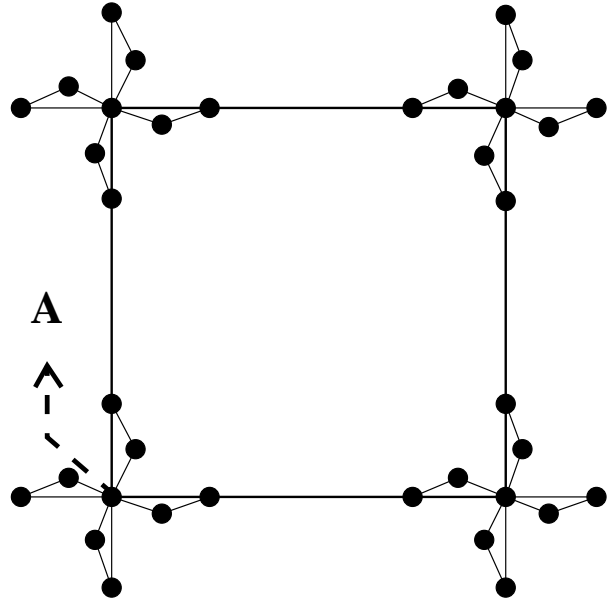}}
\caption{A decoration that leads to only the $4$-fold rotation
symmetry being preserved.} 
\label{sqspacec}
\end{figure}

We go back to the structures  in  Fig.\ \ref{sqspace} (a) and Fig.\ 
\ref{sqspace} (b).  In the  structure  (a) with the origin at A
${\bf t}_g \equiv 0$ for all $g$ and thus this point has $4mm$
symmetry.  In
Fig.\  \ref{sqspace} (b) the sites, $A$ and $B$, have $C_4$ as
their site symmetry and the mirror, $M$, acting on the object at,
say $A$,  takes it to the object at $B$. $A$ and $B$ are 
at $(\frac{1}{2}, \frac{1}{2})$ with respect to each other.
So, at $A$ and $B$ the element $M$ needs a translation
${\bf t}_M \equiv (\frac{1}{2},\frac{1}{2})$ associated with it
to make it a symmetry of the structure.  

The point group $4mm$ has eight elements.  In Fig.\  \ref{sqspace}
(b), $A$ has a site
symmetry with four elements and there are two ($\frac{8}{4} = 2$)
sites in the unit cell  with the same symmetry and connected to each
other by  the mirror operation,  $m$.  This suggests a general rule
that is fairly easy to prove. Let the point group be $P$ with $n_p$
elements. If we find a site with a local symmetry group $S$
of $n_S$ elements which is a subgroup of $P$, then there must be
$\epsilon$ such sites in the unit cell, where
\begin{equation}
\label{EPS}
\epsilon = n_p / n_s \ .
\end{equation}
These sites are {\bf equivalent} to each other, which means that
any of these sites can be
chosen as the origin for an equivalent description of the space group.
The reason for the existence of these $\epsilon$ equivalent sites is 
that there are $\epsilon$ elements in $P$ such that 
\begin{equation}
(P_1 S \cup P_2 S \cup \ldots P_{\epsilon} S )  =  P \ .
\end{equation}
If we want to construct a structure using objects with symmetry group $S$
so that it has the symmetry of the space group with point group $P$,
then each of the $\epsilon $ equivalent sites must be occupied by one
of the objects oriented such that the symmetry group $S$ is preserved
at the sites.  Obviously, the objects that can be placed at these
sites must have
\underline{at least}  the site symmetry, but they are allowed extra
symmetry. In general, this extra symmetry will not be preserved in the
actual structure. The simplest example of this is illustrated by placing
spheres on an SC lattice.  The surface of the undistorted
sphere is defined, of course, by $r= R_0$, where $R_0$ is the
radius of the sphere and $r$ is the distance from the center of the
lattice point.  When placed in the lattice the sphere will
distort and its surface will have an expansion in terms of
cubic harmonics: $r=R_0 + \epsilon ( \cos^4 \theta +
\sin^4 \theta \sin^4 \phi + \sin^4 \theta \cos^4 \phi  - 3/5)\dots$,
where $\theta$ and $\phi$ are the usual spherical angular coordinates.
The size and sign of the amplitude $\epsilon$ is determined
by the elastic response of the sphere.  We will refer back to this
concept of molecular distortion frequently.  In contrast, if
one tries to put on a site an object which does not possess all the
symmetry elements of the site, then the lattice must distort.  For
example, if the square lattice were to be decorated with,
say, a rectangle (with principal axes along the crystal axes),
then the crystal would spontaneously distort into a rectangular lattice.

\section{The Technique: Two Dimensions.}
\label{tech}

Consider the following situation in which it is assumed known that
benzene (or hexagonal) molecules are situated on a square lattice.
We will demonstrate our technique by answering the following
question: Is it possible to arrange molecules with fixed orientations
so that the unit cell contains two molecules,\cite{UC} and that the
point group of this structure is $4mm$?  For
concreteness, one may think of this molecule as having
extremely strong intramolecular bonds, so that it suffers at
most an infinitesimal distortion when placed in the crystal.
Our construction of allowed structures relies on Eq. (\ref{EPS}).
If we want to have $p$ molecules per unit cell, we look for
a site having a site symmetry group which is a subgroup of
the symmetry of the molecule and which has $n_s=n_p/p$ elements,
where we can then place the oriented molecule.  In this case
all molecules occupy sites which are equivalent in the sense mentioned
after Eq. (\ref{EPS}).  It is also possible to occupy more than
one set of equivalent sites, providing the total number of sites
in all sets of equivalents sites is $p$.  Having carried out
a construction to obtain all such sites, it will be necessary to
check a few constraints.  First of all, since we imposed a condition
on the locations of the centers of mass of the molecules,
(i. e. that the centers form a square lattice),
we must check that the proposed structures satisfy this condition.
Less trivially, we must also check that the proposed structure does
not have a higher symmetry than presupposed.  Having higher symmetry
can mean that the unit cell is smaller than desired.  (We will see this
when we later deal with octagonal molecules.) Alternatively, when
we consider the three dimensional case, we will find structures in
which the value of a structural parameter is not fixed by symmetry.
In that case, it often happens that for a special value of that
parameter additional symmetry elements appear, in which case the point
group has higher symmetry than in the generic case.

We now determine the allowed arrangements of benzene molecules.
The symmetry of benzene molecule  belongs to the point group
$6mm$. (In this notation, which is discussed in Appendix C,
$6mm$ means that the molecule has a six--fold axis of
symmetry perpendicular to its plane and it has two inequivalent
mirror planes, the normals to which lie in the plane of the
molecule.  We treat the molecule as perfectly flat and thin,
so that the additional mirror plane in the plane of the molecule
is omitted as being irrelevant to the present discussion.)
We have already discussed the $2$-d space groups with square
lattice and point group $4mm$ in the preceding discussion and
the space groups are listed in Table \ref{tsqsg}. $4mm$ has
eight elements, so if we can  find a site in the unit cell
with a site symmetry group containing four elements, there
must be two of them.  This site symmetry group of four
elements must also be a subgroup of the symmetry of the benzene
molecule ($6mm$) so that a molecule can be placed at each site
without distorting the lattice.

Accordingly, we can only place hexagonal molecules on sites
whose symmetry group is simultaneously a subgroup of that
($6mm$) of the molecule and also a subgroup of the point
group of the lattice. 
The symmetry group $2mm$, (the symmetry of a rectangle,
generators $R_2,M$, discussed in section \ref{rel})  satisfies the
requirement of being a group with four elements which is a
subgroup of both $4mm$ and $6mm$.  Thus
the problem has been transformed to that of finding sites in
the unit cell of the two space groups with $2mm$ symmetry. There are
two distinct ways of getting the subgroup $2mm$ from $4mm$.
We choose the $2$-fold rotation, $R_2$,  as one of the generators. 
For the other generator, we can choose either the mirror $M$ or the
mirror $M'$.  (Choosing $M$ and $M'$ as generators gives the
full group $4mm$.)

We have two space groups with a square lattice and point group $4mm$ 
which we consider separately in the following subsections. 

\subsection{Space group $p4mm$: ${\bf t}_{R_4} \equiv(0,0),
{\bf t}_M \equiv(0,0)$} 

From the compatibility relation, Eq.\  \ref{comp} we can find the
translations associated with all the elements, given the translations
for the generators. With the origin at point A in Fig.
\ref{sq}, ${\bf t}_{R_4} \equiv (00)$ and ${\bf t}_M \equiv(00)$.
Then the translations associated with all the elements are zero.
Thus, ${\bf t}_{R_2} \equiv (0,0)$. and ${\bf t}_{M'} \equiv (0,0)$.

The origin has $4mm$ symmetry. If the origin is shifted by
${\bf c} \equiv (c_1,c_2)$, then the new values (indicated
by primes) of the translations associated with the elements
are obtained from Eq. (\ref{oshift}) as
\begin{eqnarray}
{\bf t}_{R_4}^\prime & \equiv &  (- c_2 - c_1, c_1 - c_2) +
{\bf t}_{R_4} \equiv  (- c_2 - c_1, c_1 - c_2) \label{d1}\\
{\bf t}_{M}^\prime & \equiv &  (0 ,  - 2 c_2) +
{\bf t}_{M} \equiv  (0 ,  - 2 c_2) \label{d2} \\
{\bf t}_{R_2}^\prime & \equiv & (-2 c_1, -2 c_2) +
{\bf t}_{R_2} \equiv (-2 c_1, -2 c_2) \label{d3}\\
{\bf t}_{M'}^\prime & \equiv & (c_2 - c_1, c_1-  c_2) +
{\bf t}_{M'} \equiv (c_2 - c_1, c_1-  c_2) \ . \label{d4}  
\end{eqnarray}
We consider the two choices for $2mm$ separately. 

\subsubsection{$2mm$ with the  generators $R_2, M$.} 

The problem for this case  boils down to this, Is it possible to
choose $c_1,c_2$ such that ${\bf t}_{R_4}^\prime \not= (00)$ but
${\bf t}_{R_2}^\prime \equiv {\bf t}_M^\prime \equiv (00)$?
If the answer is yes, then we have found sites whose symmetry
group is $2mm$, with generators $R_2$, $M$.

Clearly, we may restrict $c_1$ and $c_2$ to satisfy
$0 \leq c_i < 1$.  Then to get ${\bf t}_{R_2}^\prime =0$,
we must have $c_1=0$ or $c_1=1/2$ and
$c_2=0$ or $c_2=1/2$.  To get $t_{R_4}^\prime \not= 0$
restricts us to the two possibilities: $c_1= \frac{1}{2}, \ c_2 = 0$ or
$c_1=0, \ c_2= \frac{1}{2}$.  First consider the choice
${\bf c} = (\frac{1}{2},0)$. 
This site has the desired symmetry, therefore we may place a benzene
molecule here. Since $t_M^\prime =0$, the molecule has symmetry
with respect to a mirror along the $x$--axis.  There are two choices
(which lead to equivalent structures), one of which is shown in Fig.
5a.  [The other choice is obtained by applying a 90$^{\rm o}$
rotation to the hexagon at $(\frac{1}{2},0)$.]  Having made
the selection as in Fig. 5 for the molecule at $(\frac{1}{2},0)$
the orientation of the other molecule may be determined by applying
the space group operation $(R_4, {\bf t}_{R_4}^\prime)$.  Thereby we
see that the orientation of the second molecule in the unit cell is
as shown in Fig. 5.  One can check that an equivalent structure is
obtained by the choice ${\bf c} = (0, \frac{1}{2})$.

\subsubsection{$2mm$ with  generators, $R_2,M'$}
In this case we try to make
${\bf t}_{R_2}^\prime \equiv {\bf t}_{M'}^\prime \equiv (0,0)$ and
$t_{R_4}^\prime \not= (0,0)$.  It is easy to see in this case, 
${\bf t}_{R_2}^\prime \equiv {\bf t}_{M'}^\prime \equiv (0,0)$
implies that $t_{R_4}^\prime \equiv (0,0)$.
Thus it is impossible to find a structure
satisfying our requirements.  This means that the 
structure in Fig.\  \ref{hex} (a) is the only one allowed in this
space group with 2 benzenes per unit cell. 

\subsection{Space group $p4gm$: ${\bf t}_{R_4} \equiv (0,0) , {\bf t}_{M}
\equiv ( \frac{1}{2} , \frac{1}{2} )$}
We follow the same procedures used for the previous space group,
except here we have to do a little more work to find the translations
associated with the elements of $4mm$. 

From Eq.\  (\ref{comp}) and the fact that $R_4^2 = R_2 $ we have
${\bf t}_{R_2} \equiv ( 0,0)$.  Using Eq. (\ref{comp})
and the fact that $M' = R_4 M$, we find that
${\bf t}_{M'} \equiv (\frac{1}{2}, \frac{1}{2}) $.   Both
these values pertain to the origin as shown in Fig. 1.
Thus for this space group, Eqs.\ (\ref{d2}) (\ref{d4}) become, respectively,
\begin{equation}
\label{TMP}
{\bf t}_{M}^\prime = (\frac{1}{2},\frac{1}{2}-2c_2)
\end{equation}
\begin{equation}
\label{TPMP}
{\bf t}_{M^\prime}^\prime = (\frac{1}{2}+c_2-c_1,\frac{1}{2}+c_1-c_2) \ .
\end{equation}
Since ${\bf t}_{R_4}' \equiv {\bf t}_{R_2} \equiv (0,0)$, Eqs. (\ref{d1})
and (\ref{d3}) 
also hold for this space group.

\subsubsection{$2mm$ with the  generators $R_2, M$.} 
Here we try to make ${\bf t}_{R_2}^\prime\equiv {\bf t}_M \equiv (0,0)$.
But from Eq.\ (\ref{TMP}) it is obvious that it is impossible to make
${\bf t}_M^1 \equiv (0,0)$.  Thus it is not possible to have a 
structure with two hexagons per unit cell that preserves $M$ as a
local site symmetry. 

\subsubsection{$2mm$ with the  generators $R_2, M'$.} 
We now  try  to find sites with $2mm$ symmetry using $R_2, M'$ as the
generators.  Using the relations Eq.\   \ref{d1}, \ref{d3},
and \ref{TPMP} one may verify that choosing either
${\bf c} \equiv (\frac{1}{2},0)$ or $(0,\frac{1}{2})$
makes ${\bf t}_{M'} \equiv {\bf t}_{R_2} \equiv (0,0)$
and makes ${\bf t}_{R_4} \neq (0,0) $.

The structure, shown in Fig.\  \ref{hex} (b), is constructed
for the case $c\equiv (\frac{1}{2},0)$
following the procedure outlined in the previous case.  The
other choice, ${\bf c} \equiv (0,\frac{1}{2})$, gives
the same structure rotated by $90$ degrees about the $2$-fold axis
through the origin `$O$'. 

\subsection{Inequivalent sites}

\begin{figure}
\centerline{\psfig{figure=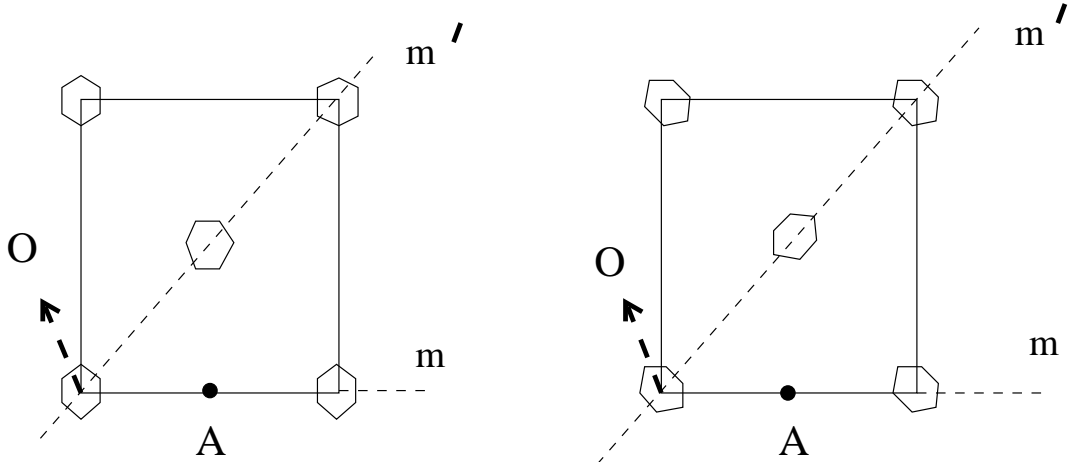}}
\caption{\label{hex} (a) shows  
hexagons in space group $p4mm$ on a square  lattice with point
group $4mm$ and  translations $\protect{\bf t}_{R_4} \equiv (0,0)$,
$\protect{\bf t}_M \equiv(0,0)$ defined with respect to origin $A$.
The choice of origin here ($O$) makes
$\protect{\bf t}_{R_4} \equiv (\protect{\frac{1}{2}},\protect{\frac{1}{2}} )
\equiv \protect{\bf t}_{M'}$, $ \protect{\bf t}_M \equiv
\protect{\bf t}_{R_2} \equiv (0 , 0)$. (b) shows 
hexagons in space group $p4gm$ for the same lattice and point group but 
the translations are  $\protect{\bf t}_{R_4} \equiv (0,0)$, 
 One of the hexagons has sides parallel to the $(1,0)$ axis and
the other has sides parallel to the $(0,1)$ axis.
 $ \protect{\bf t}_M \equiv(\protect{\frac{1}{2}}
,\protect{\frac{1}{2}}) $   defined with respect to
the origin $A$.  The choice of origin here ($O$) makes
$\protect{\bf t}_{R_4} \equiv (\protect{\frac{1}{2}},\protect{\frac{1}{2}})
\equiv \protect{\bf t}_{M}$, $ \protect{\bf t}_{M'}
\equiv \protect{\bf t}_{R_2} \equiv (0 , 0)$.}
One of the hexagons has sides parallel to the $(1,1)$ direction and
the other has sides parallel to the $(1,-1)$ direction.
\end{figure}

The procedures detailed above considered only the case when the
molecules are placed in equivalent sites, defined in Sec.
\ref{prop}. In fact, in the above example it was not possible
to place two hexagons on a square lattice on inequivalent sites.
From an algebraic point of view, that would have required finding
a group of 8 elements which was simultaneously a subgroup of the
point group of the molecule and  of the point group of the crystal.
One may also see by inspection that the above solutions do not work.
To visualize inequivalent sites, imagine  painting the
hexagons different colors, say, red and blue.  In the above
solutions, such a difference causes the $x$ and $y$ axes in
Fig. \ref{hex}a to differ and the $(1,1)$ and $(1,-1)$ directions
in Fig. \ref{hex}b to differ.
(In the first case, one axis is parallel to a mirror line
passing through the vertices of a blue hexagon, and the other axis
is parallel to a mirror line passing through the vertices of a red
hexagon.) So in this case, the lattice will distort into a rectangular
lattice.  Which axis becomes elongated depends on which
color hexagon is easier to squash.

To illustrate an arrangement involving {\it inequivalent}
sites, consider putting two octagonal molecules per unit cell
on a square lattice.
In the most general case, which we will discuss later, we will allow
the molecules to distort in response to their environment in the
crystal.  However, at first, we first consider the molecules to be
perfectly rigid.

\begin{figure}
\centerline{\psfig{figure=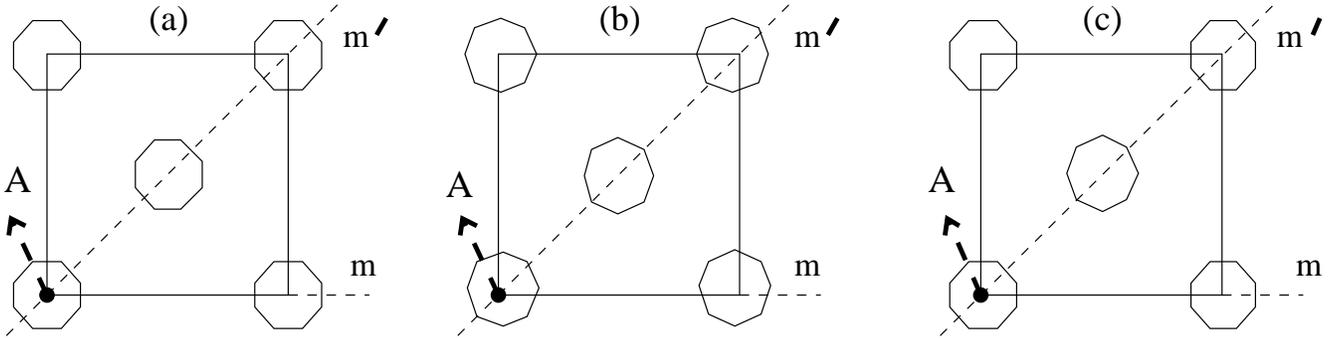}}
\caption{\label{oct1}
Placement of perfectly rigid octahedrons on a square lattice.
In panels a), b), and c) we show structures corresponding to
taking $A>0$ and $B<0$, $A<0$, $B<0$, and $A=0$, $B>0$,
respectively (See Eq. \protect\ref{POT}).  Because of the
high symmetry of the octagon,
the structures in panels a) and b) yield a square lattice
whose lattice constant is smaller than the original lattice
by $\protect\sqrt 2$.  If our aim is to obtain a diffraction pattern
in which indexing is according to the square lattice shown,
these structures are not acceptable. Structure c), which
is similar to an antiferromagnet, is acceptable.  The
effect of molecular distortions is shown in
Fig. \protect\ref{nonrigid} and \protect\ref{nonqnz}.}
\end{figure}

We start by looking for arrangements of rigid octagons on equivalent
sites on a square lattice.  For this purpose
we follow the procedure we used to place oriented hexagons: we
look for sites which have a symmetry
group having four elements and which is a subgroup of both  $4mm$
(the point group of the lattice) and $8mm$ (the point group of the molecule).
Following exactly the analysis for benzene, we have to determine
the orientations of octagons which have mirror planes along the
$x$ and $y$--directions.  There are two such orientations:
one, used in panel (a), in which the mirror planes cut the faces
at their midpoints and (b) another in which the mirrors pass through
a vertex.  Using this information we obtain the structures
shown in panels (a) and (b) of Fig. \ref{oct1}.  But since these
structures index to a smaller unit cell, we reject them.\cite{UC}
Note that, as in Fig. \ref{hex}, the orientations of the two
molecules were assumed to be related by a symmetry operation.

Now we consider the possibility of occupying two inequivalent
sites in the unit cell with oriented octagons.  In this case
we look for sites which have a symmetry group which is a subgroup
of both $4mm$ and $8mm$ which contains eight elements.
We will need two
such sites, because according to Eq. (\ref{EPS}) each site in this
case has only a single representative.  In this case the two
sites are not related to each other via symmetry.  For the site symmetry
group to be the full $(4mm)$ point group, we need to have that
${\bf t}_g \equiv 0$ for all $g$. This is only possible for the
space group $p4mm$.  Inspection of Eqs. (\ref{d1}) -- (\ref{d4})
shows that the sites having this high symmetry are ${\bf c} = (0,0)$
and ${\bf c} = ( \frac{1}{2}, \frac{1}{2})$, as one could have
intuited.  On each of these sites we may independently choose
orientations of the octagon which are invariant under the
operations of $4mm$.  That fixes the octagon to be oriented so
that a line from its center to either a vertex or to the center
of a face is parallel to a nearest neighbor direction.
We thereby arrive at the structures shown in Fig. \ref{oct1}.
However, the choices where the two sites are actually
equivalent was found previously.  They correspond to a unit cell
with only one octagon and were therefore rejected.  The choice (c)
of Fig. \ref{oct1} does correspond to two inequivalent octagons
in the unit cell and is therefore is the desired structure.

We wish to emphasize that there is no physical requirement that
identical molecules should occupy crystallographically
equivalent sites.  So we now turn to a brief discussion
of orientational potentials which give rise to a structure
such as (c) of Fig. \ref{oct1} which involve inequivalent sites.
For simplicity this discussion is carried out for a two dimensional
system, but the conclusions hold for three dimensional systems also.
To obtain the symmetry of the orientational potential, it is convenient
to imagine the molecules interacting via an atom--atom potential,
$\Phi(I,J)$, where $I$ and $J$ label molecules and
\begin{equation}
\Phi(I,J) = \sum_{i\in I} \sum_{j\in J} \phi({\bf R}_i -
{\bf R}_j) \ ,
\end{equation}
where $i$ and $j$ label atoms, $i \in I$ indicates that atom $i$
is in molecule $I$, and $\phi$ is a generic atom--atom potential.
To discuss the symmetry of $\Phi(I,J)$ we consider expanding it
in powers of ${\bf r}_i \equiv {\bf R}_i - {\bf R}_I$, where
${\bf R}_I$ is the position of the center of molecule $I$.
Atomic coordinates will appear in the combinations
\begin{equation}
r_i^m e^{i m \phi_i} \biggl( 1 + a_2 r_i^2 + a_4 r_i^4  + \dots \biggr)
\ ,
\end{equation}
where $r_j e^{i\phi_j} = x_j + i y_j$ and $a_{2n}$ depends on the details of
the potential $\phi$.  To discuss questions of symmetry, we may ignore
terms involving $a_2$, $a_4$, etc.  Then
\begin{equation}
\label{EXPAND}
\Phi(I,J) = \sum_{M=-\infty}^\infty \sum_{N=-\infty}^\infty
a(M,N) \sigma_M (I) \sigma_N(J) e^{-i (M+N) \phi_{IJ}} \ ,
\end{equation}
the coefficients $a(M,N)$ depend on $R_{I,J}$ and
$R_{IJ}e^{i \phi_{IJ}} = X_{IJ} + i Y_{IJ}$, where
${\bf R}_{IJ} = {\bf R}_I - {\bf R}_J$, etc.
Also
\begin{equation}
\sigma_M(I) = \sum_{i \in I} r_i^{|M|} e^{iM\phi_i} \ .
\end{equation}
For completely rigid octagons $\sigma_M$ is only nonzero when
$M$ is divisible by 8.  If we specify the orientation of an octagon
by the angle, $\theta$, the line joining its center to a vertex makes with a
nearest neighbor direction, then one sees that
$\sigma_M \propto e^{iM \theta}$.
For our purposes, it suffices to consider the following intermolecular
potential for {\it rigid} octagons:
\begin{equation}
\label{POT}
V ( \{ \theta_i \} )  = A \sum_i V_0 \cos (8 \theta_i)
+ B \sum_{\langle ij \rangle} \cos (8 \theta_i ) \cos (8 \theta_j) \ ,
\end{equation}
where $\langle ij \rangle$ indicates that the sum is over pairs
of nearest neighbors.  Then the structures in Fig. \ref{oct1}
correspond to choices of signs of the parameters in the potential
as indicated in the figure caption.
Within the approximation of completely rigid molecules there is
thus only one way to arrange octagons on a square lattice
(see panel c of Fig. \ref{oct1})  so
that they belong to a space group having ($4mm$) point group
symmetry and have two molecules per unit cell.  This example
is important because it shows explicitly that there is no
requirement that two identical molecules should remain
equivalent when placed in the crystal.  To reach this
case merely requires that $B>0$ in Eq. \ref{POT} and that
$A$ plays no role in determining the structure ($A=0$, for example).

Now let us see what happens when we relax the assumption that
the molecules are perfectly rigid.  We therefore consider the
limit when all the intramolecular force constants are arbitrarily
large, but are not actually infinite.  It is clear that in this
limit the $I$th molecule will distort in response to the crystal field,
$V_c$, of the square lattice.  To see this analytically, we look at
terms in the multipole expansion of Eq. (\ref{EXPAND}) with $M=4$ and
$N=0$.  Denoting this term for molecule $I$ by $V_c^{(4)}(I)$, we have
\begin{equation}
V_c^{(4)} (I) = A \sum_J \sum_{i \in I} r_i^4 \cos (4 \phi_i - 4 \phi_{IJ} ),
\end{equation}
where the constant $A$ depends on the details of the potential.
Considering only nearest neighbor interactions, we may write
\begin{equation}
V_c^{(4)}(I) = A \sum_{i \in I} r_i^4 \cos (4 \phi_i ) .
\end{equation}
Now we allow for a distortion of the molecule by setting
\begin{equation}
\label{DISTORT}
r_m = r [ 1 + (-1)^m \epsilon ] \ , \ \ \ \ \ 
\phi_m = \theta + ( m \pi /4) - (-1)^m \delta \ ,
\end{equation}
where the atoms in the octagon are numbered sequentially from
1 to 8.  To leading order in $\epsilon$ and $\delta$ we have
\begin{equation}
\label{LINEAR}
V_c^{(4)}(I) = B_\epsilon \epsilon \cos (4 \theta) 
+ B_\delta \delta \sin (4 \theta) \ ,
\end{equation}
where the $B$'s are constant dependent on the details of the potential.
Since the intramolecular interactions give rise to a energy for such
distortions of the form
\begin{equation}
E = \frac{1}{2} C_\epsilon \epsilon^2 + \frac {1}{2} C_\delta \delta^2 \ ,
\end{equation}
it is clear that in the generic case,  (i. e.  barring
an accidental vanishing of $B_\epsilon$ or $B_\delta$)
the molecule will be distorted
in the crystal.  The distortions are illustrated for the cases
$\theta=0$ and $\theta=\pi/4$ in Fig. 7.

In contrast, one may ask whether it is possible for the
structures of panels (a) and (b) of Fig. \ref{oct1}, which
we rejected as giving rise to a smaller unit cell,
to lead to acceptable structures when given inequivalent distortions,
as in Fig. \ref{nonqnz}.  Of course, such a structure is possible.
But it involves an additional removal of degeneracy.  Analytically,
the situation is as follows.  We have already seen the
existence of terms in the energy of the form
\begin{equation}
E = \sum_i \biggl( \frac{1}{2} C_\epsilon \epsilon(I)^2 + B_\epsilon
\epsilon(I) + \frac {1}{2} C_\delta \delta(I)^2 + B_\delta \delta (I)
\biggr) \ , \end{equation}
where $\delta(I)$ and $\epsilon(I)$ are the distortions in molecule $I$.
To get inequivalent distortions
(As in the right panel of Fig. \ref{oct1})
we need to invoke the term
with $M=\pm 4$, $N=\pm 4$ from Eq. (\ref{EXPAND}), which gives rise to
an energy involving $\epsilon(I)$ of the form
\begin{equation}
\label{BREAK}
E = - D_\epsilon \sum_{<IJ>} \epsilon(I) \epsilon(J) \ ,
\end{equation}
where $D_\epsilon$ is a constant.
As the temperature is lowered to reach the orientationally
ordered phase, one may ask about the role of this interaction.
Clearly $D_\epsilon$ has to exceed a critical strength in order for
there to be an instability towards $\epsilon(I)$ becoming nonzero
at nonzero wavevector, before orientational ordering of octagons
occurs.  In other words, since this requires $D_\epsilon/C_\epsilon$
to exceed a critical threshold value, this is not the generic case.

The situation is only simple in the limit when the molecules
are nearly infinitely rigid.  In this limit, the only
distortions which are allowed are those which involve
linear coupling, as in Eq. (\ref{LINEAR}).  If one
considers such cooperative distortions, then, at least
in principle, one should allow for arbitrary rearrangements
of the atoms, even rearrangements totally unrelated to the
structure of the original molecule.  In that case, essentially
anything is possible.  So the only meaningful question to ask
is how do nearly perfectly rigid molecules orient within a given
space group or class of space groups.  The group theoretical
method described here enables a treatment of perfectly rigid molecules.
Then to take nonrigidity into account, one can allow distortion due
to linear interactions with the crystal field as outlined above.

\section{The Technique: Three Dimensions}
\label{3d}

We will now discuss arrangements in $3$-dimensions using the example
of  icosahedral C$_{60}$ molecules (symmetry group : $Y_h$) 
in a cubic lattice. 
At high temperatures it is known that the molecules are on an FCC
lattice. Each site on this lattice is occupied by a  molecule
that is  randomly reorienting and hence each site has the full
symmetry of the FCC lattice. At about $250$K there is a first order
phase transition to an SC lattice.\cite{PAHPRL}  Evidence from
NMR shows that the molecules are no longer randomly reorienting. The obvious
conclusion is that the molecules are now orientationally ordered and 
there are four molecules per simple cubic unit cell, one each at
[$(000),({1 \over 2},{1 \over 2},0),({1 \over 2},0,{1 \over 2}) \ 
\text{and} \ (0,{1 \over 2},{1 \over 2})$]. 

Using our  technique, we will now find all simple cubic structures
that allow four icosahedral  molecules per unit cell at the FCC sites.
(It will turn out that for space groups with inversion
symmetry, the only way to have four molecules
per unit cell it to have them at the sites given above.)  Out of
the possible structures, the actual one must be determined
by a structurally sensitive experiment, such as a scattering
experiment.

We need to note here some definitions and terminology.
First we define various point group operations in Table \ref{t3dimage}
and we illustrate these for the cube in Fig. \ref{cube}
A non-axial point group is one
which has more than one $n$-fold  axis of symmetry ($n > 2$), while
axial point groups have a unique $n$-fold axis of symmetry. 
In $3$-dimensions there are only $5$ nonaxial crystallographic 
point groups, $O_h \ldots T$ listed in Table \ref{nonaxial} and
discussed in appendix \ref{anonaxial}   
The rest are all axial point groups which are of the  seven kinds
listed in Table \ref{axial} and whose properties are described briefly in
appendix \ref{aaxial}.  Finally, to obtain all possible simple cubic
space groups having oriented icoshedra on FCC lattice sites, we
list, in Table \ref{tscsg}, the translation vectors associated
with the generators for all simple cubic space groups.  For
completeness, we also give, in Tables \ref{tfcsg} and \ref{tbcsg},
the analogous data for FCC and body-centered cubic space groups.

Concerning the placement of icosahedra there are two scenarios:
we either find all four sites equivalent to
each other or we find sites that are inequivalent, in which case the
molecules at different sites are  not related to each other through
some symmetry operation. The inequivalent sites can be partitioned as
$(1+3)$, where there is one site with the full symmetry of the point
group and three other sites equivalent to each other, $(2+2)$,
$(1+1+2)$ and $(1+1+1+1)$. 

\subsection{Icosahedral symmetry}

The isolated C$_{60}$ molecule (bucky ball) forms
a truncated icosahedron. The symmetry of the
structure is the same as that of an icosahedron, $Y_h$, which  is the
highest possible non-axial symmetry group,  with $60$ elements. Fig.\
\ref{c601} shows a icosahedron and its relation to the symmetries of a
cube.  It is
not a crystallographic symmetry group. The crystallographic subgroups
of $Y_h$ can be classified under three classes, as can be seen from Fig.\
\ref{c601}, 

1) $T_h$ and its subgroups. (generators = $\overline{r}_3$ and $r'_2$.)

2) $\overline{3} m$ and its subgroups. (generators = $\overline{r}_3$
and a mirror passing through a vertex of the icosahedron and
containing the $3$-fold rotoinversion axis.)

3) $mmm$ and its subgroups.  (generators = the mirrors along the
$xy$,$yz$ and $zx$ planes.)

In what follows we use the tables \ref{nonaxial}, \ref{axial}
and \ref{t3dimage}. 

\subsection{Simple Cubic $O_h$ space groups.}

We first consider $4$ equivalent sites.  Since $O_h$ has $48$ elements
we need to find a subgroup of $Y_h$ and $O_h$ with $12$
elements. There are two possibilities, $T$ and $\overline{3}m$. 

$T$ has the generators $r_3,r'_2$. Also we can see that $(e + r_4 +
\overline{r}_4 + i) T = O_h$. Thus we need to have the translations
associated with $r_3,r'_2$ be zero while the translations associated
with $r_4, \overline{r}_4, i$ have to be non-zero so that the site has
$T$ symmetry.

We will now present arguments to prove that it is impossible to have
sites with only $T$ symmetry in the $O_h$ simple cubic space
groups. Of course, this can be easily verified by going through the
algebra.

Let us assume there exist four sites in a unit cell with $T$
symmetry. Consider a tetrahedron  at the origin. The three other
sites must have tetrahedra related to the one at the origin by the 
the operations $r_4$, $\overline{r}_4$ and  $i$ respectively.  

Symmetry elements of $T$, applied at the origin,  must be  symmetries  of
the whole crystal.  This means they should either leave the other
three molecules invariant   or exchange their sites and orientations. 
The rotation about the $3$-fold along the $111$ axis leaves all
the molecules invariant which means, to leave their positions
invariant they must all lie on the $111$ axis, but applying the
$3$-fold along say $\bar{1}11$ would want the molecules to lie along
 this axis, which would not be compatible. So we have  a proof by
contradiction.

Let us now consider $\overline{3}m$ with generators
$\overline{r}_3,m''$. Arguments given above do not 
prevent the occurrence of sites with  $\overline{3}m$ symmetry and they
do occur as shown below. 
 We see that $(e + r_4 + r_4^2 +
r_4^3) \overline{3}m = O_h$. So we need to find a space group that has
${\bf t}_{\overline{r}_3} \equiv (000) \equiv {\bf t}_{m''}$ and ${\bf t}_{r_4}$
nonzero. From $m'' = r'_2 m r'_2$ and $m = r_4 \overline{r}_3$ and the
compatibility relations, we get for the four $O_h$ simple cubic space
groups listed in Table \ref{tscsg},  
\begin{eqnarray}
{\bf t}_{r_4} &  =  &  \left(\begin{array} {c}
	(0,0,0)       \\
	(0,0,{1 \over 2})     \\
	({1 \over 2},{1 \over 2},0)   \\
	({1 \over 2},{1 \over 2},{1 \over 2})
\end{array} \right) \ ; \ 
{\bf t}_{m''}  =  \left( \begin{array} {c}
	(0,0,0)       \\
	(\frac{1}{2},\frac{1}{2},{1 \over 2})     \\
	(0,0,0)   \\
	({1 \over 2},{1 \over 2},{1 \over 2})
\end{array} \right) 
\end{eqnarray} 
Since all the space groups already have  ${\bf t}_{\bar{r}_3} \equiv (000)$
we are restricted to ${\bf c}  \equiv ({1 \over 2},{1 \over 2},{1 \over 2})$.
This will not change any of the ${\bf t}_g$,  since $g c - c $ for any 
$g \ \epsilon  \ O_h$ is zero. So we need to find a
space group that has the required symmetry at the origin.

The only suitable  space group with $\bar{3}m$ as site symmetry 
is the one with ${\bf t}_{\bar{r}_3} \equiv (000), {\bf t}_{r_4} \equiv
({1 \over 2},{1 \over 2},0)$. This is space group $Pn \overline 3 m$.
In this structure a C$_{60}$ molecule is placed at the origin with one
of its $\overline{3}$ axis along the $111$ axis and rotated about this
axis till the mirror of the molecule containing the $3$-fold axis  is
along the mirror $m''$ of the lattice. Applying
the operation $r_4$ to this creates a molecule pointing in the
$(11\overline{1})$ direction and since
${\bf t}_{r_4} \equiv ({1 \over 2},{1 \over 2},0)$, the
molecule is placed at $({1 \over 2},{1 \over 2},0)$. 
The other two molecules in the unit cell
can be placed by noting that repeated application of the $3$-fold
rotation about the $(111)$ direction should be a symmetry and hence
there are two more molecules at $(0,{1 \over 2},{1 \over 2})$ pointing in the
$(\overline{1}11)$ direction and at $({1 \over 2},0,{1 \over 2})$
pointing in the $(1\overline{1}1)$ direction.

{\bf Inequivalent sites}

The $(1+3)$, $(1+1+2)$ and $(1+1+1+1)$ partitions are 
 easily excluded since the unique site must
have $O_h$ symmetry which is not a symmetry of the molecule under
consideration. If we restrict attention to  the FCC sites then $(2+2)$
is also excluded since the two equivalent sites must necessarily lie on
the $(111)$ axis and be separated by
$(\frac{1}{2},\frac{1}{2},\frac{1}{2})$ if the cubic symmetry of the
lattice is to be preserved. This reasoning will exclude the
consideration of this partition from all the following point  groups. 
Actually we need only consider the $(1+3)$  partition in the following
examples. 

\subsection{Simple Cubic $T_h$ space groups.}

$T_h$ has $24$ elements. For four equivalent sites  the 
only subgroup of $Y_h$ and $T_h$ with $6$ elements is  
$\bar{3}$, whose generator is  $\bar{r}_3$. 

Since all the $T_h$ simple cubic space groups are listed with
${\bf t}_{\overline{r}_3} \equiv (000)$, we are again restricted to
${\bf c} \equiv ({1 \over 2},{1 \over 2},{1 \over 2})$ which leaves
all the translations
unaltered. Hence we are again reduced to finding a space group
whose origin has the required symmetry.  All space groups with 
${\bf t}_{\overline{r}_3} \equiv (000)$ and  non-zero  ${\bf t}_{m'}$
can be used and hence there are two possibilities, 

1) ${\bf t}_{\overline{r}_3} \equiv (000), {\bf t}_{r_2'}
\equiv ({1 \over 2},0,{1 \over 2})$ [$Pn\overline{3}$]

This structure is identical to the one found in the $O_h$ case
above except for the fact that the molecule can be rotated an
arbitrary angle about the local $3$-fold. If this angle is such
that a mirror plane of the molecule coincides with a crystal (110)
direction, then the crystal has an additional mirror symmetry
and the structure should be classified under the higher symmetry
$O_h$ space group, which we found already. 

2) ${\bf t}_{\overline{r}_3} \equiv (000), {\bf t}_{r_2'}
\equiv ({1 \over 2},{1 \over 2},0)$ [$Pa\overline{3}$]

By arguments similar to those used above we find that four molecules
are at 

i) $(000)$ with a three-fold axis in the $(111)$ direction. 

ii) $({1 \over 2},{1 \over 2},0)$ with a three-fold axis in the
$(\overline{1}1\overline{1})$ direction. 

iii) $({1 \over 2},0,{1 \over 2})$ with a three-fold axis in the
$(1\overline{1}\overline{1})$ direction.

iv)  $(0,{1 \over 2},{1 \over 2})$ with a three-fold axis in the
$(\overline{1}\overline{1}1)$ direction.

Applying the $2$-fold rotation about the $110$ axis gives us the other
possible realization of this space group with ${\bf t}_{\overline{r}_3}
\equiv (000), {\bf t}_{r_2'} \equiv (0,{1 \over 2},{1 \over 2})$. 

{\bf Inequivalent Sites} : We have already discussed why we need
consider only the $(1+3)$ partition. The unique site must have $T_h$
symmetry, which automatically restricts us to considering only the
space group with ${\bf t}_{r'_2} \equiv (000)$ since all the other space
groups with $T_h$ point group cannot have a site with full $T_h$
symmetry.  The $3$ equivalent sites must have $mmm$ ($8$ elements)
symmetry with the translation associated with the $3$-fold rotation
nonzero at these sites. Let the three mirrors along the $xy$,$yz$ and 
$zx$ planes be
labeled $m'_2,m'_3,m'$. The translations associated with all three of
them is zero at the origin. Under an origin translation we have
$\Delta {\bf t}_{m'} \equiv (0,-2 c_2,0)$, $\Delta {\bf t}_{m'_2} \equiv
(0,0,-2 c_3)$, 
$\Delta {\bf t}_{m'_3} \equiv (-2 c_1,0,0)$ and 
$\Delta {\bf t}_{r_3} \equiv (c_3-c_1,c_1-c_2,c_2-c_3)$. 
There are two choices, the first of which is
${\bf c} \equiv (\frac{1}{2},0,0)$ and its cyclic permutations,
all of which lead to icosahedra being placed at the edge centers with
mirrors along the mirrors of the lattice.  Since the molecules do not
form an FCC lattice, we reject this choice.  The second choice is
${\bf c} \equiv (\frac{1}{2},\frac{1}{2},0)$ and its cyclic permutations,
which lead to icosahedra placed at the face centers.  This choice,
which yields a structure of space group Pm$\overline 3$, does satisfy
our requirements. 

\subsection{Other Simple Cubic Space Groups.}

The other cubic point groups, $O$, $T$, and $T_d$, lack inversion
symmetry.  Since we have required the molecules to be placed on
an FCC lattice, allowed structures must have inversion symmetry.
(In a moment, we will discuss the extent to which this restriction can be
relaxed.)  So, unless we allow a molecular distortion which
removes inversion symmetry, there are no allowable structures
of these cubic point groups.

Now we address the question as to whether inversion symmetry should
actually be required.  We argue that removal of inversion symmetry
requires a breaking of symmetry analogous to that we discussed in
conjunction with Eq. (\ref{BREAK}).  If we place molecules on an FCC
lattice, there is no generic loss of inversion symmetry.
For nearly infinitely rigid molecules, a loss of inversion
symmetry is not expected.  When one considers the related problem of
alkali metal doped C$_{60}$, the situation is different.  There
one has a reason why generically inversion symmetry could be
broken, and consideration of space groups without inversion symmetry
is reasonable.\cite{TYPRL}  

We now discuss briefly, without derivation, structures which can
occur if we relax the condition that the molecule has inversion symmetry.
First we consider the case when all four sites are equivalent.
Each of the structures found requiring inversion symmetry has a
counterpart without inversion symmetry, as one would expect, since one can
easily envision a small perturbation which breaks inversion symmetry.
Thus the Pn$\overline 3$m structure of the $O_h$ point group becomes
the P$4_232$ structure of $O$ point group symmetry.  In this case,
a loss of inversion symmetry does not cause any displacement of
the molecules from their FCC lattice sites.  In the case of
the Pn$\overline 3$ and Pa$\overline 3$ structures, loss of inversion
symmetry implies a displacement of the molecules away from their
positions when there is inversion symmetry.  (Of course, this is
the generic case.  One can imagine the accidental case when this
displacement is zero.)  In both cases, this displacement is such that the
center of mass of the molecule is translated along the three-fold axis.
Thus the analog of Pn$\overline 3$ is P23 and that of Pa$\overline 3$
is P$2_13$.
Presumably, if this displacement is small enough, it could
be consistent with the experimental determination that the centers
of mass of the molecules form an FCC lattice.  In any event, the
loss of inversion symmetry is NOT generic and requires either
noninfinitesimal distortion of the molecule or an external perturbation,
like doping.

When we consider inequivalent sites, we find that the only
possibility occurs within $T$ point group symmetry.
Starting from Pm$\overline 3$, one can remove inversion symmetry.
In this case, the molecules are not displaced from their
position in the presence of inversion symmetry.  But, as in the
case of equivalent sites, this removal of inversion symmetry
does require either a noninfinitesimal distortion of the molecule,
or a perturbation to break inversion symmetry,
and thus is NOT generic.

\section{Conclusions}
\label{SUMMARY}

We have demonstrated a constructive algorithm for finding
orientations of molecules whose centers of mass are at or
infinitesimally near specified locations.  In this algorithm
one obtains candidate structures from which
one eliminates structures which are unacceptable because they
do not have molecules at
the specified locations.  Also, in classifying structures one
has to take care to identify additional symmetries that may be introduced
in the construction, especially when structural parameters whose
values are not fixed by symmetry, are assigned special values.
In addition, we have given an extensive discussion of the role
of symmetry breaking in which the molecule spontaneously distorts
or its center of mass spontaneously is displaced from the assigned
position.  Unless there is some external perturbation (like doping)
such spontaneous symmetry breaking is not the generic case which
applies when the ratio of force constants associated with 
intramolecular bonds to those associated with intermolecular
interactions is arbitrarily large.  If such spontaneous
symmetry breaking is allowed, the problem becomes poorly
formulated in that molecular and atomic positions arbitrarily
far from those desired become permissible.

{\bf Acknowledgements} R.S. thanks the NSF for a
pre--doctoral fellowship.  This work was supported in part by the
NSF under Grants No. 91--22784 and
DMR88-19885 of the MRL program.  ABH also thanks the USIEF
for support via a Fulbright Fellowship and the Department of
Physics and Astronomy for their hospitality.
We also thank Y. Ohrn, R. Lifshitz, 
T. C. Lubensky, and P. A. Heiney for useful discussions. We
acknowledge useful correspondence with T. Hahn.

\appendix
\section{Derivation of the Compatibility relation.}
\label{acomp}

Let $g$, $h$,  $k =gh$ etc. be the elements of the point group of a
periodic structure. Let the symmetries of  the structure be
$(g,{\bf t}_g)$, $(h,{\bf t}_h)$, $(k,{\bf t}_k)$ etc.
The compound operation, $(h,{\bf t}_h)$ followed by
$(g, {\bf t}_g)$, is a symmetry of the structure.  Thus, 
\begin{eqnarray}
\label{comp1}
(g(h,{\bf t}_h), {\bf t}_g) & = & gh + g {\bf t}_h + {\bf t}_g 
\nonumber \\ & = & k + g {\bf t}_h + {\bf t}_g .
\end{eqnarray}
But $(k,{\bf t}_k)$ is a symmetry of the structure, so comparing
this with Eq.\ \ref{comp1} gives, 
\begin{equation}
{\bf t}_k = {\bf t}_{gh} = g {\bf t}_h + {\bf t}_g . 
\end{equation}

\section{Derivation of the Origin Shift Relation.}
\label{aoshift}

Let $(g,{\bf t}_g)$ be a symmetry of  a periodic structure.
If we apply this symmetry operation to the structure then the
coordinates ${\bf x}$ of a point $A$ in the  structure transforms
to ${\bf x}'$ given by 
\begin{equation}
\label{shift1}   
{\bf x}' = g {\bf x} + {\bf t}_g .
\end{equation}
In a new coordinate system, which is obtained by translating the
original system  by $\bf c$, the coordinates of $A$ are given by
${\bf x}_n$, where 
\begin{equation}
{\bf x}_n = {\bf x} - {\bf c} . 
\end{equation}
Using this relation in Eq.\ \ref{shift1} we get, 
\begin{equation}
{\bf x}'_n + {\bf c} = g ({\bf x}_n + {\bf c}) + {\bf t}_g 
\end{equation}
which gives us
\begin{equation}
{\bf x}'_n = g {\bf x}_n  + {\bf t}_g + g {\bf c} - {\bf c} . 
\end{equation}
This means in the new frame, 
\begin{equation}
({\bf t}_g)_{\text{new frame}} =
({\bf t}_g)_{\text{old frame}} + \Delta {\bf t}_g , 
\end{equation}
where
\begin{equation}
\Delta {\bf t}_g = g {\bf c} - {\bf c} . 
\end{equation}

\section{Axial Point Groups}
\label{aaxial}

Axial point groups are those with a unique axis of high symmetry. 
Obviously the symmetry groups of the five regular Platonic solids
(cube, tetrahedron etc.) don't qualify as axial point groups. In this
paper an {\bf $n$-fold rotation} is $r_n$, that is,
a rotation by $\frac{2\pi}{n}$ radians. The {\bf $n$-fold
rotoinversion} is $\overline{r}_n$, which is defined by, 
\begin{equation}
\overline{r}_n = r_n i, 
\end{equation}
where $i$ is the {\bf inversion}. A {\bf dihedral axis}, $d$, is a
$2$-fold rotation axis perpendicular to the high
symmetry axis. $m$ is a {\bf mirror containing} the high
symmetry axis. $h$ is a {\bf mirror perpendicular}
to the high symmetry axis.  The following are a few
properties of these point groups. 

i) Since 
\begin{equation}
r_2 h = i, 
\end{equation}
if any two of the three elements exist the third also exists. 

ii) If two mirrors $m_1$ and $m_2$ exist at an angle $\frac{\pi}{n}$
with respect to each other, then the line of intersection of the
mirrors is an $r_n$ axis. There will be $n$ mirrors in such a point
group, which are in two distinct conjugate classes if $n$ is even and
in the same class if $n$ is odd. 
Two elements $g_1,g_2$ of a point
group belong to the same conjugate class if $h g_1 h^{-1} = g_2$, where
$h$ is also an element of the point group. A special case 
is when the two mirrors are perpendicular to each other in
which case the line of intersection is $r_2$. Obviously, if a mirror
contains an $r_n$ axis, then there are $n$ mirrors which all intersect
at $r_n$ and are at an angle $\frac{\pi}{n}$ to each other. The new
mirrors can be generated using 
\begin{equation}
r_n m = m'. 
\end{equation}
Similar statements can be made for dihedrals, in which case
the line perpendicular to the axes at the intersection point
of the two dihedrals forms the $r_n$ axis. 

iii) $\overline{r}_n$ exists, in which case we can have three cases,

\noindent
a) If $n = 2 p +1$, then $\overline{r}_n^n = i$, $\overline{r}_n^{n+1} = r_n$
and $\overline{r}_n^{2n} = e$. 

\noindent
b) If $n=2 p$; $p$ is odd, then $\overline{r}_n^p = h$, $\overline{r}_n^2 =
r_p$ and $\overline{r}_n^n =e$. 

\noindent
c) If $n=2 p$; $p$ is even, then $\overline{r}_n^p = r_2$, $\overline{r}_n^2 =
r_p$ and $\overline{r}_n^n =e$.

iv) $\overline{r}_n$ and $m$ exist. In this case, 
\begin{equation}
\overline{r}_n m = i r_n m = i m' = d', 
\end{equation}
which means that there always exists a dihedral axis perpendicular to
each mirror. Thus this case is the same as $\overline{r}_n$ occuring with
$d$. Again three cases to be considered separately following the
discussion of the previous case, 

\noindent
a) If $n = 2 p +1$, then, since $r_n$ and $i$ exist we have $n$
mirrors at $\frac{\pi}{n}$ from each other and $n$ dihedrals
bisecting the angle between the mirrors.  

\noindent
b) If $n=2 p$; $p$ is odd, then $r_p$ and $h$ exist and thus there are
$p$ mirrors are $\frac{\pi}{p}$ with respect to each other and $p$
dihedrals parallel to $m$. 

\noindent
c) If $n=2 p$; $p$ is even, then only $r_p$ exists, no $h$ or $i$ and
thus there are $p$ mirrors at $\frac{\pi}{p}$ with respect to each
other and $p$ dihedrals bisecting the angle between the mirrors. 

\underline{Notations} : Two conventions of naming point groups are
prevalent,  Schoenflies and International. Tables \ref{nonaxial} and
\ref{axial} list all the kinds of point groups in both notations. 

In  the {\bf Schoenflies} convention, capital letters are used
to designate the class of the point group and subscripts of small
letters and
numbers are used to specify the point group completely. For example
in $D_{4d}$, $D$ says that the point group is a dihedral group, that
is it has dihedral axes, the $4$ says it has a $4$-fold axis and $d$
says that it has diagonal mirrors, that is the mirrors bisect the
angle between the dihedrals. Another example is $C_{4h}$,  where the
point group has a $4$-fold axis and a horizontal mirror, that is the
mirror is perpendicular to the $r_4$ axis.

In the {\bf International}
convention,  small letters and numbers
specify the point group. For example  $2mm$ means the point group has
a $2$-fold axis and  the two $m$'s specify that there are two classes
of mirrors. $4/mmm$ means that there is a $4$-fold axis and three
classes of mirrors with one of them perpendicular to the $4$-fold
axis. 
 
\section{Non-Axial Point Groups}
\label{anonaxial}

There are only five non-axial crystallographic point groups 
$O_h$, $T_h$, $O$, $T_d$ and $T$. They are related to each other as
follows, $T \subset T_d \subset O_h$,  $T \subset T_h \subset O_h$ and
$T \subset O \subset O_h$. $(e + \overline{r}_4) T = T_d$, $(e + r_4)
T_d = O_h$, $(e + r_4) T = O$, $(e + i) O = O_h$,
  $(e + i) T = T_h$ and  $(e + r_4) T_h = O_h$.  
  The main axial subgroups of each of these
groups are  listed below, with their generators given in brackets. 

\begin{itemize}

\item
$T$  -  $3 (r_3)$, $222 (r'_2 \text{and its partners along the} x
\text{and} z \text{axis}$). 

\item
$T_d$  -  $2mm (m,m'')$, $\overline{4}2m (m,r'_2)$, $\overline{4} (\overline{r}_4)$.

\item
$O$  -  $422 (r_2,r'_2)$, $222 (r_2,r''_2)$, $32 (r_3,r''_2)$.

\item
$T_h$  -  $\overline{3} (\overline{r}_3)$, $mmm$ (mirrors along the
coordinate planes). 

\item
$O_h$  -  $\overline{3}m (\overline{r}_3,m'')$, $4/mmm (i,m,m')$.

\end{itemize}

\begin{center}
\begin{table}
\caption{Images of $(x, y) $ due to various symmetry operations of
the square lattice ($p4mm$).} 
\label{t2dimage}
\end{table} 
\begin{tabular}{|c|c|c|c|c|c|}
\hline
\ Operation \  & \  $e$ \ & \ $m$ \ & \  $m'=r_4 m $ \ & \ $r_2$
\ & \ $  r_4$ \   \\
\hline
Image & $ x y  $ 
&$ x \overline{y}  $ 
&$ y x   $ 
&$  \overline{x} \overline{y}  $ 
&$ \overline{y} x  $ \\
\hline
\end{tabular}
\end{center}
\begin{center}
\begin{table}
\caption{The space groups with square lattice and point group $4mm$.
The origin is assumed to be at a lattice point (point A in Fig. 1).}
\label{tsqsg}
\end{table}
\begin{tabular}{|c||c|c|c|}
\hline
\ Point group\ & \  generators \ &  \  ITC number \ & \ translations \  \\
\hline
$4mm$ & $r_4,m $  & 
$ 11 (p4mm) $  &  ${\bf t}_{r_4} \equiv ( 0, 0 ),
{\bf t}_{m} \equiv ( 0, 0 ) $ \\
\cline{3-4}
& & $ 12 (p4gm) $ & ${\bf t}_{r_4} \equiv ( 0, 0 ), {\bf t}_{m} \equiv
(\frac{1}{2},\frac{1}{2})$  \\ 
\hline
\end{tabular}
\end{center}
\begin{center}
\begin{table}
\caption{Images of $(x, y, z) $ due to various symmetry operations of the
cube.}
\label{t3dimage}
\end{table}
\begin{tabular}{|c|c|c|c|c|c|c|c|c|c|c|c|}
\hline
\ Operation  \ & \ $e$ \ & \ $i$ \  & \ $m =$ \ & \ $r_2=$ \ & \ $m'$ \
&\  $r_2' = $ \ 
&  \ $r_4$ \ & \ $r_3$ \ & \ $r_3'=$ \ & $m''=$ \ & \ $r_2''=$ \ \\ 
  &  & & $r_4 \overline{r}_3$ & $im$ & & $r_4^2$
&  & & $r_3 r_2'$  & $r_2' m r_2' $ & $r_2' r_2 r_2'$ \\ 
\hline
Image & $ x y z $ 
&$ \overline{x} \overline{y} \overline{z}  $ 
&$ \overline{y} \overline{x} z  $ 
&$  y  x \overline{z}  $ 
&$ x \overline{y} z  $ 
&$ \overline{x} y \overline{z}  $ 
&$ z y \overline{x}  $
&$ z x y  $
&$ \overline{z} \overline{x} y  $
&$ y x z   $ 
&$ \overline{y} \overline{x} \overline{z}   $ 
\\
\hline
\end{tabular}
\end{center}

\begin{center}
\begin{table}
\centering
\caption{The Non-Axial  Point  Groups.}
\label{nonaxial}
\end{table}
\begin{tabular}{|l||l|l|l|l|}
\hline
& Generators & Schoenflies  & \# of   &   Generating
\\
 & & International & elements &   relations \\
\hline
$1$ &  $ r_3, r'_2 $                   & $T$  & $12$ &
$ r_3^3= (r')^2 = (r_3 r'_2)^3=e $\\
    &                & $23$            &      & 
$ r_3 r'_2 = r'_3$\\
\hline
$2$ &  $ r_3,{{\overline r}_4}(=i r_4)$& $T_d$ & $24$ & 
$r_3^3={{\overline r}_4}^4=({{\overline r}_4} r_3 )^2=e,$\\  
    &                & $\overline{4}3m$&      & 
$ \overline{r}_4 r_3 = m, \overline{r}_4^2 = r'_2$ \\
\hline
$3$ & $ r_3, r_4 $                     & $ O$ & $24$ &
$ r_3^3=r_4^4=(r_4 r_3)^2=e,$\\
    &                                  & $432$&      &
$ r_4 r_3 = r_2 $ \\
\hline
$4$ & $ {\overline r}_3,r'_2(= r_4^2)$  & $T_h$& $24$ &
 ${\overline r}_3^6=(r'_2)^2=({\overline r}_3^3 r'_2 )^2=e,$\\ 
    &                                  & $\frac{2}{m}\overline{3}$ & &
$\overline{r}_3^3 =i ;  i r'_2 =r'_2 i = m' $\\
\hline
$5$ & $ {\overline r}_3 (= i r_3),r_4$ &$O_h$ & $48$ & 
${\overline r}_3^6=r_4^4=(r_4 {\overline r}_3)^2=e,$ \\
    &                                  &$ \frac{4}{m} \overline3
\frac{2}{m} $ & &
$ r_4 \overline{r}_3 =m ; 
{\overline r}_3^3 r_4= r_4 {\overline r}_3^3$\\
\hline
\end{tabular}
\end{center}

\begin{center}
\begin{table}
\centering
\caption{The Axial Point Groups.$^a$}
\label{axial}
\end{table}
\begin{tabular}{|l|l|l|l|l|l|}
\hline
& Generators & Schoenflies    & \# of   &   Generating   &
Crystallographic \\
&            &(International) & elements&   relations    & Realizations  \\
\hline
$1$& $r_n$   &$C_n(n)$        & $n$     & $r_n^n = e$    & $n=1,2,3,4,6$ \\
\hline \hline
$2$&$\overline{r}_n$ & $S_{2n}(\overline{n}); n = 2 p + 1$ & $2n$ & 
$\overline{r}_n^n =i, \overline{r}_n^{2n}=e$ & $n=1,3$ \\ 
\cline{3-6}
   &                 & $C_{ph}(\overline{n}); n=2p; p$ odd & $n$  &
$\overline{r}_n^2=r_p,\overline{r}_n^p = h$  & $n=2,6$ \\
\cline{3-6}
   &                 & $S_n(\overline{n}); n=2p; p$ even &   $n$ &
$\overline{r}_n^n=e,\overline{r}_n^p = r_2$  & $S_4 (\overline{4})$ \\
\hline \hline
$3$&$r_n,d$          & $D_n (n2); n = 2p+1$& $2n$ & $r_n^n=e=d^2$ &
$D_3 (32)$ \\
\cline{3-3} \cline{5-6}
   &                 & $D_n(n22); n =2p$  &      & $r_n^p =r_2;d^2 =e$
& $n=2,4,6$ \\
\hline \hline 
$4$&$r_n,m$          & $C_{nv} (nm); n = 2p+1$& $2n$ & $r_n^n=e=m^2$ &
$C_{3v} (3m)$ \\
\cline{3-3} \cline{5-6}
   &                 & $C_{nv}(nmm); n =2p$  &      & $r_n^p =r_2;m^2 =e$
& $n=2,4,6$ \\
\hline \hline 
$5$&$r_n,h$          & $C_{nh} (\overline{2n}); n = 2p+1$& $2n$ &
$r_n^n=e=h^2$ & $n=1,3$ \\
\cline{3-3} \cline{6-6}
   &                 & $C_{nh}(n/m); n =2p$  &      & & 
$n=2,4,6$ \\
\hline \hline 
$6$&$r_n,m,h$   & $D_{nh} (\overline{2n}2m); n = 2p+1$& $4n$ & $r_n^n=e=h^2
= m^2$ & $D_{3h} (\overline{6}2m)$ \\
\cline{3-3} \cline{5-6}
   &                 & $D_{nh}(n/mmm); n =2p$  &      & $r_n^p
=r_2;m^2 =e=h^2$ & $n=2,4,6$ \\
\hline \hline 
$7$&$\overline{r}_n,m$ & $D_{nd}(\overline{n}m); n = 2 p + 1$ & $4n$ & 
$\overline{r}_n^n =i, \overline{r}_n^{2n}=e=m^2$ & $D_{3d}(\overline{3}m)$ \\ 
\cline{3-6}
   &                 & $D_{ph}(\overline{n}2m); n=2p; p$ odd & $2n$  &
$\overline{r}_n^2=r_p,\overline{r}_n^p = h$  & $D_{3h} (\overline{6}2m)$ \\
\cline{3-6}
   &                 & $D_{pd}(\overline{n}2m); n=2p; p$ even &   $2n$ &
$\overline{r}_n^n=e,\overline{r}_n^p = r_2$  & $D_{2d}(\overline{4}2m)$ \\
\hline 
\end{tabular}

\vspace{0.3 cm}
\noindent
a)   Note that $C_{1h}(\overline 2)$, $C_{3h} (\overline 6)$,
and $D_{3h} (\overline 6 2m)$ each appear twice in this listing, also
technically, $r_n$ with $n=2$ should be classified separately. 
\end{center}

\begin{center}
\begin{table}
\caption{The Simple Cubic Space Groups.}
\label{tscsg}
\end{table}
\begin{tabular}{|c||c|c|c|}
\hline
Point group &  generators  &  ITC number  &  translations    \\
\hline
$O_h$ & $\overline{r}_3, r_4 $  & 
$ 221 (Pm\overline{3}m)$  &  ${\bf t}_{\overline{r}_3} \equiv (0,0,0),
{\bf t}_{r_4} \equiv (0,0,0) $\\
\cline{3-4}
& & $ 222 (Pn\overline{3}n)$ & ${\bf t}_{\overline{r}_3} \equiv
(0,0,0), {\bf t}_{r_4} \equiv (0,0,\frac{1}{2}) $\\
\cline{3-4}
& & $ 224 (Pn\overline{3}m) $ & ${\bf t}_{\overline{r}_3} \equiv
(0,0,0), {\bf t}_{r_4} \equiv (\frac{1}{2},\frac{1}{2},0) $ \\
\cline{3-4}
& & $ 223 (Pm\overline{3}n)$ & ${\bf t}_{\overline{r}_3} \equiv
(0,0,0), {\bf t}_{r_4} \equiv (\frac{1}{2},\frac{1}{2},\frac{1}{2}) $ \\
\hline
$T_h$& $\overline{r}_3, r_2' $ 
& $ 200 (Pm\overline{3})$ & ${\bf t}_{\overline{r}_3} \equiv
(0,0,0), {\bf t}_{r_2'} \equiv (0,0,0) $ \\
\cline{3-4}
& & $ 201 (Pn\overline{3}) $ & ${\bf t}_{\overline{r}_3} \equiv
(0,0,0), {\bf t}_{r_2'} \equiv (\frac{1}{2},0,\frac{1}{2}) $ \\
\cline{3-4}
& & $ 205 (Pa\overline{3}) $ & ${\bf t}_{\overline{r}_3} \equiv
(0,0,0), {\bf t}_{r_2'} \equiv (\frac{1}{2},\frac{1}{2},0) $ \\
\cline{4-4}
& &  & ${\bf t}_{\overline{r}_3} \equiv
(0,0,0), {\bf t}_{r_2'} \equiv (0,\frac{1}{2},\frac{1}{2}) $ \\
\hline
$O$& $r_3, r_4 $ 
& $ 207 (P432)$ & ${\bf t}_{r_3} \equiv
(0,0,0), {\bf t}_{r_4} \equiv (0,0,0) $ \\
\cline{3-4}
& & $ 213 (P4_132)$ & ${\bf t}_{r_3} \equiv
(0,0,0), {\bf t}_{r_4} \equiv (\frac{3}{4},\frac{1}{4},0) $ \\
\cline{3-4}
& & $ 208 (P4_232)$ & ${\bf t}_{r_3} \equiv
(0,0,0), {\bf t}_{r_4} \equiv (\frac{1}{2},\frac{1}{2},0) $ \\
\cline{3-4}
& & $ 212 (P4_332)$ & ${\bf t}_{r_3} \equiv
(0,0,0), {\bf t}_{r_4} \equiv (\frac{1}{4},\frac{3}{4},0) $ \\
\hline
$T_d$& $r_3, \overline{r}_4 $ 
& $ 215 (P\overline{4}3m) $ & ${\bf t}_{r_3} \equiv
(0,0,0), {\bf t}_{\overline{r}_4} \equiv (0,0,0) $ \\
\cline{3-4}
& & $ 218 (P\overline{4} 3n)$ & ${\bf t}_{r_3} \equiv
(0,0,0), {\bf t}_{\overline{r}_4} \equiv (0,0,\frac{1}{2}) $ \\
\hline
$T$& $r_3, r_2' $ 
& $ 195 (P23)$ & ${\bf t}_{r_3} \equiv
(0,0,0), {\bf t}_{r_2'} \equiv (0,0,0) $ \\
\cline{3-4}
& & $ 198 (P2_13)$ & ${\bf t}_{r_3} \equiv
(0,0,0), {\bf t}_{r_2'} \equiv (\frac{1}{2},\frac{1}{2},0) $ \\
\hline
\end{tabular}
\end{center}

\begin{center}
\begin{table}
\caption{The Face-Centered  Cubic Space Groups listed using  the same
bases as in the simple cubic space groups. Thus, for every site in this
list add three more  at $(0,\protect\frac{1}{2},\protect\frac{1}{2})$
and its cyclic permutations. Note that translations by these vectors
are now identities.
This allows the use of table \protect\ref{t3dimage} without any
changes. 
 The natural bases for this lattice are ${\bf a_1}
= \protect\frac{1}{2}(0,1,1)$ and its cyclic permutations.}
\label{tfcsg}
\end{table}
\begin{tabular}{|c||c|c|c|}
\hline
 Point group &  generators  &  ITC number  &  translations    \\
\hline
$O_h$ & $\overline{r}_3, r_4 $  & 
$ 225 (F 4/m\overline{3}2/m)$  &  ${\bf t}_{\overline{r}_3} \equiv
(0,0,0), {\bf t}_{r_4} \equiv (0,0,0) $\\
\cline{3-4}
 & & $ 226 (F4/m\overline{3}2/c)$ & ${\bf t}_{\overline{r}_3} \equiv
(0,0,0), {\bf t}_{r_4} \equiv (0,0,\frac{1}{2}) $\\
\cline{3-4}
& & $ 227 (F4_1/d\overline{3}2/m) $ & ${\bf t}_{\overline{r}_3} \equiv
(0,0,0), {\bf t}_{r_4} \equiv (\frac{3}{4},\frac{3}{4},0) $ \\
\cline{4-4}
& & & ${\bf t}_{\overline{r}_3} \equiv
(\frac{1}{4},\frac{1}{4},\frac{1}{4}), {\bf t}_{r_4}
 \equiv (\frac{1}{4},\frac{1}{4},\frac{1}{4}) $ \\
\cline{3-4}
& & $ 228 (F4_1/d\overline{3}2/c)$ & ${\bf t}_{\overline{r}_3} \equiv
(0,0,0), {\bf t}_{r_4} \equiv (\frac{3}{4},\frac{1}{4},0) $ \\
\cline{4-4}
& & & ${\bf t}_{\overline{r}_3} \equiv
(\frac{1}{4},\frac{1}{4},\frac{3}{4}),
 {\bf t}_{r_4} \equiv (\frac{1}{4},\frac{1}{4},\frac{1}{4}) $ \\
\hline
$T_h$& $\overline{r}_3, r_2' $ 
& $ 202 (F2/m\overline{3})$ & ${\bf t}_{\overline{r}_3} \equiv
(0,0,0), {\bf t}_{r_2'} \equiv (0,0,0) $ \\
\cline{3-4}
& & $ 203 (F2/d\overline{3}) $ & ${\bf t}_{\overline{r}_3} \equiv
(0,0,0), {\bf t}_{r_2'} \equiv (\frac{1}{4},0,\frac{1}{4}) $ \\
\cline{4-4}
& &  & ${\bf t}_{\overline{r}_3} \equiv
(\frac{1}{4},\frac{1}{4},\frac{1}{4}), {\bf t}_{r_2'} \equiv (0,0,0) $ \\
\hline
$O$& $r_3, r_4 $ 
& $ 209 (F432)$ & ${\bf t}_{r_3} \equiv
(0,0,0), {\bf t}_{r_4} \equiv (0,0,0) $ \\
\cline{3-4}
& & $ 210 (F4_132)$ & ${\bf t}_{r_3} \equiv
(0,0,0), {\bf t}_{r_4} \equiv (\frac{1}{4},\frac{1}{4},\frac{1}{4}) $ \\
\hline
$T_d$& $r_3, \overline{r}_4 $ 
& $ 216 (F\overline{4}3m) $ & ${\bf t}_{r_3} \equiv
(0,0,0), {\bf t}_{\overline{r}_4} \equiv (0,0,0) $ \\
\cline{3-4}
& & $ 219 (F\overline{4} 3c)$ & ${\bf t}_{r_3} \equiv
(0,0,0), {\bf t}_{\overline{r}_4} \equiv (0,0,\frac{1}{2}) $ \\
\hline
$T$& $r_3, r_2' $ 
& $ 196 (F23)$ & ${\bf t}_{r_3} \equiv
(0,0,0), {\bf t}_{r_2'} \equiv (0,0,0) $ \\
\hline
\end{tabular}
\end{center}

\begin{center}
\begin{table}
\caption{The Body-Centered  Cubic Space Groups listed using  the same
bases as in the simple cubic space groups. Thus, for every site in this
list add one  more  at
$(\protect\frac{1}{2},\protect\frac{1}{2},\protect\frac{1}{2})$. 
 Note that translations by this  vector is 
 now  an identity. 
This allows the use of table \protect\ref{t3dimage} without any
changes. 
The natural bases for this lattice are ${\bf a_1}
= \protect\frac{1}{2}(\overline{1},1,1)$ and its cyclic permutations.}
\label{tbcsg}
\end{table}
\begin{tabular}{|c||c|c|c|}
\hline
 Point group &  generators  &  ITC number  &  translations    \\
\hline
$O_h$ & $\overline{r}_3, r_4 $  & 
$ 229 (I4/m\overline{3}2/m)$  &
${\bf t}_{\overline{r}_3} \equiv (0,0,0), {\bf t}_{r_4} \equiv (0,0,0) $\\
\cline{3-4}
 & & $ 230 (I4_1/a\overline{3}2/d)$ & ${\bf t}_{\overline{r}_3} \equiv
(0,0,0), {\bf t}_{r_4} \equiv (\frac{1}{4},\frac{3}{4},\frac{3}{4}) $\\
\hline
$T_h$ & $\overline{r}_3, r_2' $ 
& $ 204 (I2/m\overline{3})$ & ${\bf t}_{\overline{r}_3} \equiv
(0,0,0), {\bf t}_{r_2'} \equiv (0,0,0) $ \\
\cline{3-4}
& & $ 206 (I2_1/a\overline{3}) $ & ${\bf t}_{\overline{r}_3} \equiv
(0,0,0), {\bf t}_{r_2'} \equiv (\frac{1}{2},0,0) $ \\
\hline
$O$& $r_3, r_4 $ 
& $ 211 (I432)$ & ${\bf t}_{r_3} \equiv
(0,0,0), {\bf t}_{r_4} \equiv (0,0,0) $ \\
\cline{3-4}
& & $ 214 (I4_132)$ & ${\bf t}_{r_3} \equiv
(0,0,0), {\bf t}_{r_4} \equiv (\frac{1}{4},\frac{3}{4},\frac{3}{4})$ \\
\hline
$T_d$& $r_3, \overline{r}_4 $ 
& $ 217 (I\overline{4}3m) $ & ${\bf t}_{r_3} \equiv
(0,0,0), {\bf t}_{\overline{r}_4} \equiv (0,0,0) $ \\
\cline{3-4}
& & $ 220 (I\overline{4} 3d)$ & ${\bf t}_{r_3} \equiv
(0,0,0), {\bf t}_{\overline{r}_4} \equiv
 (\frac{1}{4},\frac{3}{4},\frac{3}{4}) $ \\
\hline
$T$& $r_3, r_2' $ 
& $ 197 (I23)$ & ${\bf t}_{r_3} \equiv
(0,0,0), {\bf t}_{r_2'} \equiv (0,0,0) $ \\
\cline{3-4}
& & $ 199 (I2_13)$ & ${\bf t}_{r_3} \equiv
(0,0,0), {\bf t}_{r_2'} \equiv (0,\frac{1}{2},\frac{1}{2}) $ \\
\hline
\end{tabular}
\end{center}

\begin{center}
\begin{table}
\centering
\caption{The Space  Groups for $4$ C$_{60}$ Molecules at equivalent FCC
Sites.}
\label{spequiv}
\end{table}
\begin{tabular}{|c||c|c|c|c|}
\hline
&Space group & Site & FCC & Inversion \\  
\hline
$1$ &  $ O_h ({\bf t}_{r_4} \equiv ( {1 \over 2}, {1 \over 2}, 0),
Pn\bar{3}m  $  & $\overline{3} m$ & $ Y $ & $ Y $ \\
\hline
$2$ &  $ T_h ({\bf t}_{r'_2} \equiv ( {1 \over 2},0, {1 \over 2}),
Pn\bar{3} $  & $\overline{3} $ & $ Y $ & $ Y $ \\
\hline
$3$ &  $ T_h ({\bf t}_{r'_2} \equiv ( {1 \over 2}, {1 \over 2}, 0) ,
 Pa\bar{3}$  & $\overline{3} $ & $ Y $ & $ Y $ \\
\hline
$4$ &  $ O ({\bf t}_{r_4} \equiv ( {1 \over 2}, {1 \over 2}, 0),
P4_232  $  & $32$ & $ Y $ & $ N $ \\
\hline
$5$ &  $ T ({\bf t}_{r'_2} \equiv ( 0 , 0, 0),P23 $  &
$3$ & $ N^{a}$ & $ N $\\
\hline
$6$ &  $ T ({\bf t}_{r'_2} \equiv ({1 \over 2}, {1 \over 2},0),P2_13 $  &
$3$ & $ N^{a}$ & $ N $ \\
\hline
\end{tabular}

\noindent
a)  N means that, as discussed in the text, in the generic case
the structure is not FCC, but involves a nonzero displacement of the
molecules such that the lattice is still SC, but the molecule do
not quite sit on FCC lattice sites.

\end{center}

\begin{center}
\begin{table}
\centering
\caption{Space Groups for $4$ C$_{60}$ Molecules at Inequivalent FCC
Sites.}
\label{spinequiv}
\end{table}
\begin{tabular}{|c||c|c|c|c|}
\hline
&Space group & Site & FCC & Inversion \\
\hline
$1$ &  $ T_h ({\bf t}_{r'_2} \equiv ( 0,0,0), Pm\bar{3} $  &
$T_h + 3(mmm) $ & $ Y $ & $ Y $ \\
\hline
$2$ &  $ T ({\bf t}_{r'_2} \equiv (0, 0, 0), P23 $  &
$T + 3(222)$ & $ Y $ & $ N $ \\
\hline
\end{tabular}
\end{center}


\begin{figure}
\centerline{\psfig{figure=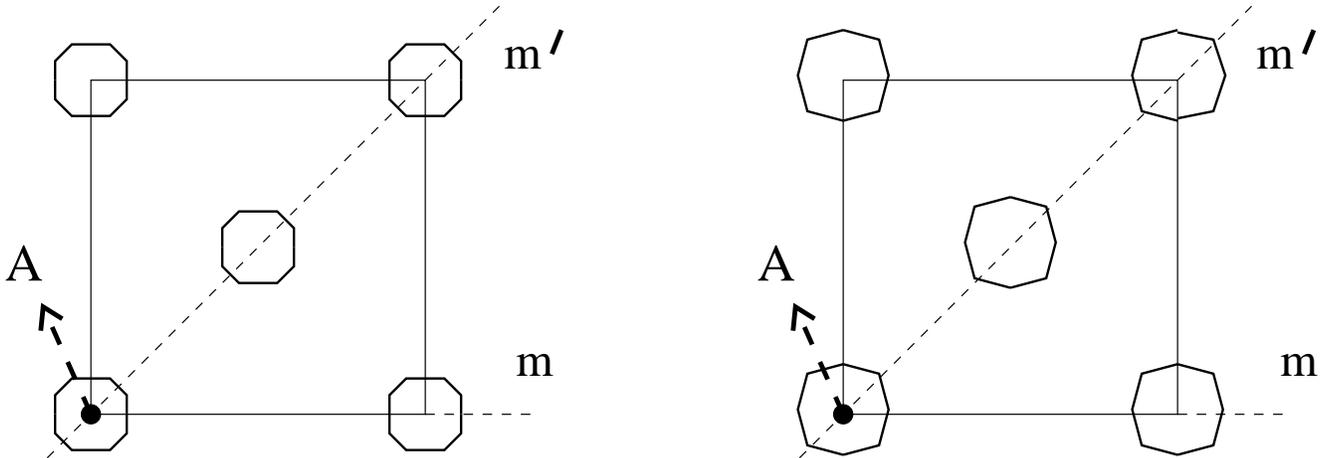}}
\caption{\label{nonrigid}
Placement of nonrigid octahedrons on a square lattice. The
left--hand structure  corresponds to $A>0$ and $B<0$ in
Eq. (\protect\ref{POT}) and the right--hand one to
$A<0$ and $B<0$.  These are the structures one must obtain
in the limit when the molecules are almost perfectly rigid.
(In this limit one can imagine the molecule held together
by springs whose force constants are arbitrarily large.)
Whether, compared to the free molecule,
the nearest--neighboring faces
(vertex angles) becomes smaller, as shown, or larger depends on
the nature of the intermolecular interactions.
The left panel corresponds to Eq. (\protect\ref{DISTORT}) with
$\epsilon=0$ and $\delta \sin (4 \theta) > 0$ and the right panel to
$\delta=0$ and $\epsilon \cos (4 \theta) > 0$.
In both cases all molecules remain equivalent
and the unit cell is smaller by $\protect\sqrt 2$ than that shown.}
\end{figure}


\begin{figure}
\centerline{\psfig{figure=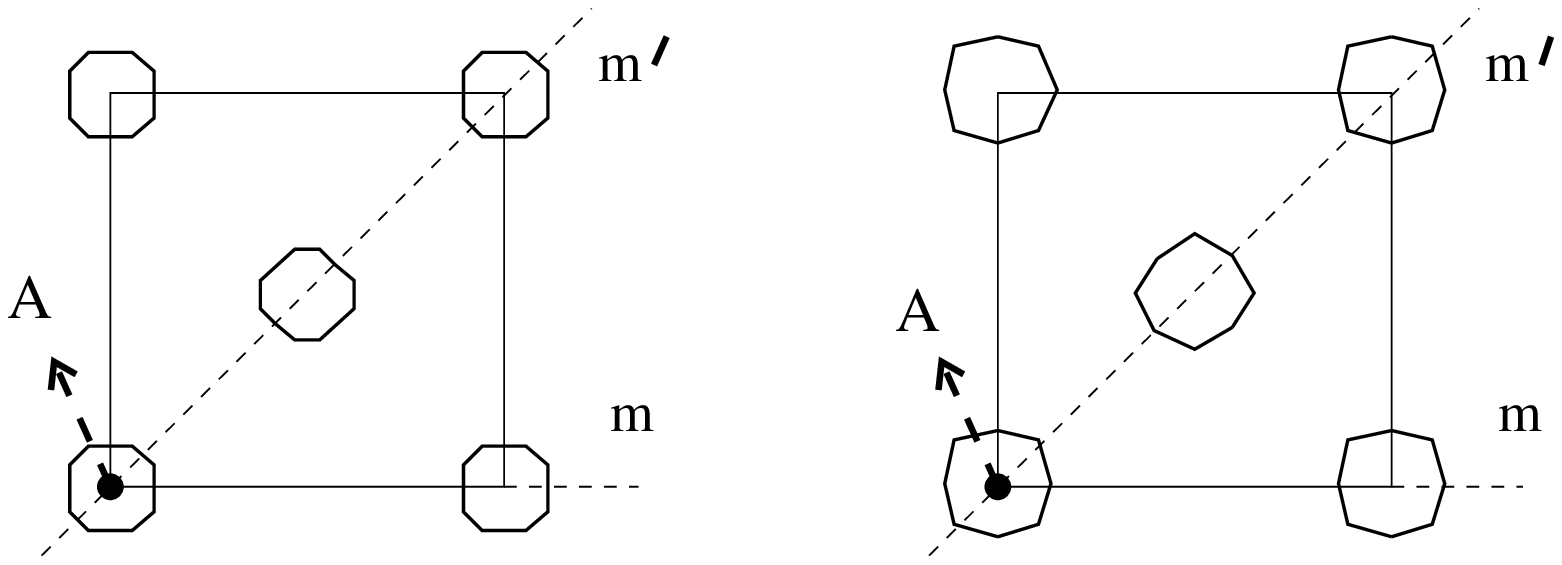}}
\caption{\label{nonqnz}
Placement of nonrigid octahedrons on a square lattice. The
left--hand structure  corresponds to
$A>0$ and $B<0$ in
Eq. (\protect\ref{POT}) and the right--hand one to
$A<0$ and $B<0$.  In this case the molecules are inequivalent
by reason of having different distortions.  However, to reach
this state the spring constants of the molecule can not
be arbitrarily large, but have to be smaller than some
critical value dependent on the intermolecular interactions.}
\end{figure}


\begin{figure}
\centerline{\psfig{figure=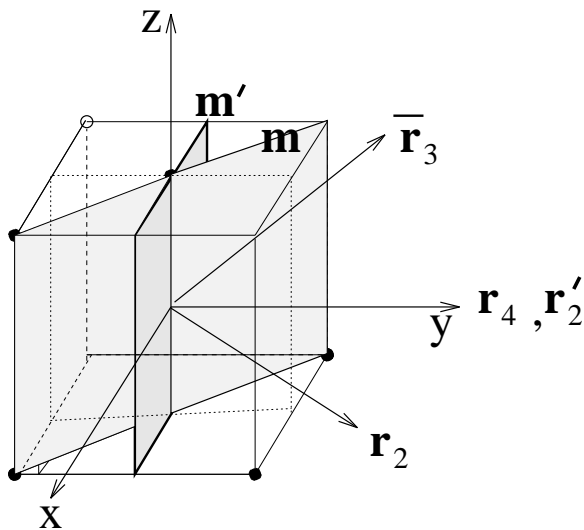,width=3.in}}
\caption{The symmetries of a cube.}
\label{cube}
\end{figure}


\begin{figure}
\centerline{\psfig{figure=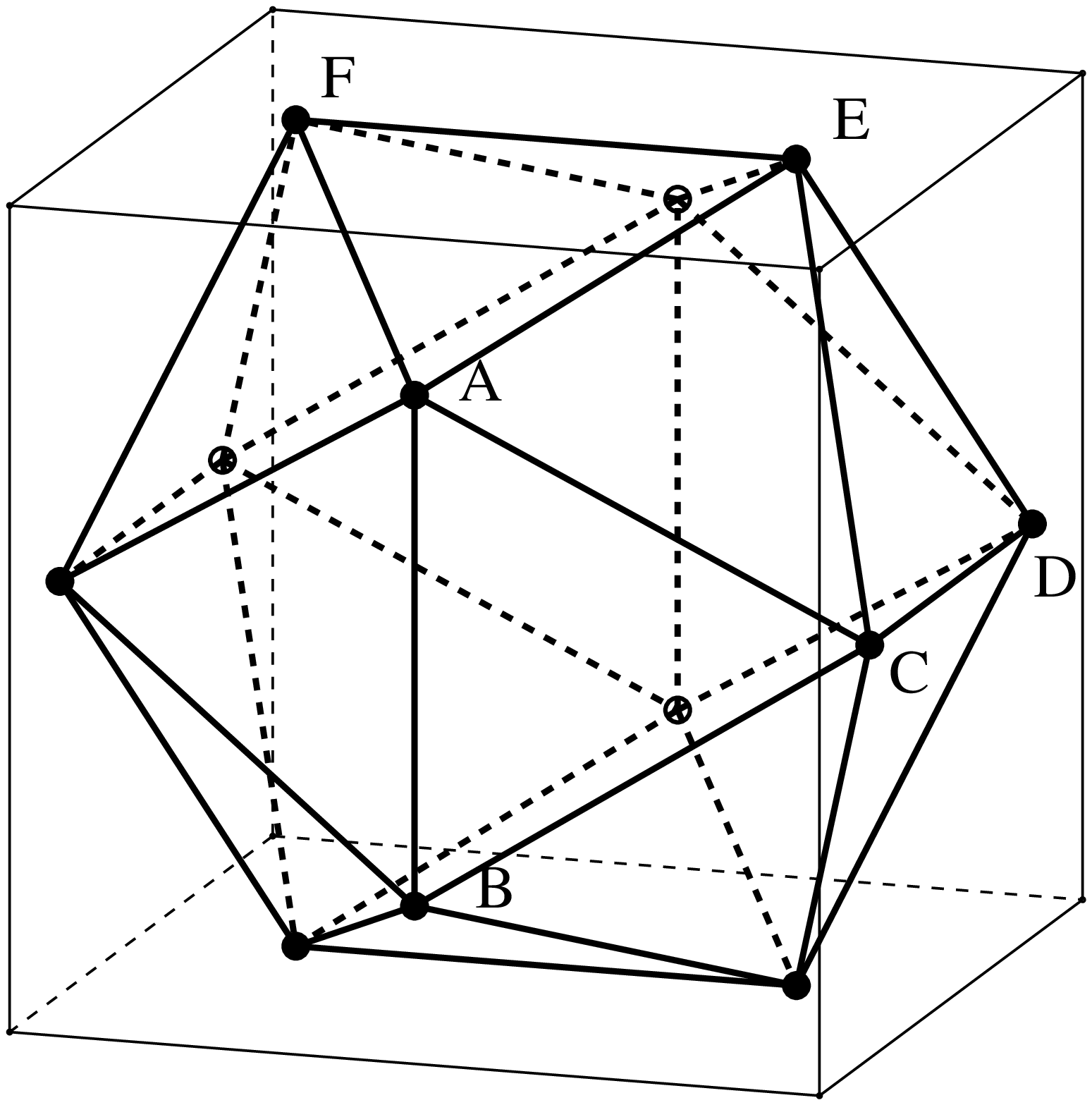,width=3.in}}
\caption{An Icosahedron bounded 
by  a cube. The edges $AB$ $CD$ etc. are on the face of the cube and
the ratio of the length of the edges of the icosahedron and the edge
of the cube is the golden mean. Each vertex of the icosahedron is  a
$5$-fold axis of symmetry.}
\label{c601} 
\end{figure}


\begin{thebibliography}{99}
\bibitem{UC}
As discussed in the introduction, we have in mind a situation where
the elastic diffraction spectrum is indexed for a square lattice
(of side $a$) containing two molecules.  We assume that the
presence of superlattice reflections (relative to the orientationally
disordered phase) precludes having one molecule per unit cell with
a lattice constant $a/\sqrt 2$.
\end{thebibliography}
\end{document}